\documentclass{aa}
\newcommand{\mcol}[3]{\multicolumn{#1}{#2}{#3} }
\newcommand{\struut}{\rule[-2ex]{0ex}{5.2ex}}
\newcommand{\struutup}{\rule{0ex}{3.2ex}}
\newcommand{\struutdown}{\rule[-2ex]{0ex}{2ex}}
\usepackage{graphicx}

\begin{document}

\title{Multicolour CCD Measurements of Visual Double and Multiple Stars. III\thanks{Based 
on observations collected at the National Astronomical Observatory, Rozhen, and the Astronomical 
Observatory, Belogradchik, both operated by the Institute of Astronomy, Bulgarian Academy of 
Sciences. Also based on data obtained by the Hipparcos astrometry satellite.},\thanks{Appendix 
A and Tables~4-6 are only available in electronic form, respectively at http://www.edpsciences.org 
(App. A), or via anonymous ftp to cdsarc.u-strasbg.fr or at http://csdweb.u-strasbg.fr/cgi-bin/qcat?/A+A/XXX/YYY (Tab.~4-6).} }

\author{P. Lampens \inst{1}  
\and A. Strigachev \inst{2}
\and D. Duval \inst{1}}

\offprints{P. Lampens}

\institute{Royal Observatory of Belgium, Ringlaan 3, B-1180 Brussels, Belgium \\
  \email{Patricia.Lampens@oma.be}
\and Institute of Astronomy, Bulgarian Academy of Sciences, 72 Tsarigradsko 
     Shosse Blvd., 1784 Sofia, Bulgaria \\
  \email{anton@astro.bas.bg} }
\date{Received \dots; accepted \dots}

\titlerunning{CCD Measurements of Double and Multiple Stars}
\authorrunning{P.~Lampens et al.}

\abstract
{Recent CCD observations were performed in the period 1998-2004 for a large sample of visual double
and multiple stars selected from the Hipparcos Catalogue and/or from the Gliese Catalogue of Nearby
Stars.}
{Accurate astrometric and photometric data allowing to characterize the individual components are 
provided. These data are confronted to Hipparcos data or to data from an older epoch in order to
assess the nature of the observed systems.} 
{We simultaneously apply a Moffat-Lorentz profile with a similar shape to all detected components
and adjust the profile parameters from which we obtain the relative astrometric position (epoch,
position angle, angular separation) as well as differential multi-colour  photometry (filters (B)VRI).} 
{We thus acquired recent data for 71 visual systems of which 6 are orbital binaries, 27 are nearby and
30 are multiple systems. In three cases, the systems remained unresolved. 23 new components were 
detected and measured. Two new visual double stars of intermediate separation were also found. 
The estimated accuracies in relative position are 0.04$\degr$ and 0.01$\arcsec$ respectively, 
while those in differential photometry are of the order of 0.01-0.02 mag in general.} 
{The nature of the association of 55 systems is evaluated. New basic binary properties are derived 
for 20 bound systems. Component colours and masses are provided for two orbital binaries.}


\maketitle

\section{Introduction}
The context of the present work is the field of visual binaries (double) and multiple stars.  We report
on the acquisition of recent astrometric and multi-passband photometric data  following the procedure
described in our previous work (\cite{la02} (Paper~I), \cite{st} (Paper~II)). During the years
1998-2004, we performed regular CCD observations of a sample of visual double and multiple stars
selected from the Gliese Catalogue of Nearby Stars and/or from the Hipparcos Catalogue (\cite{hc}) at
the National Observatory of Rozhen (NAO) and  at the Astronomical Observatory of Belogradchik (AOB),
both situated in Bulgaria. Our goal  is to improve the knowledge of the distributions of the true
separations, relative motions,  (total and individual) masses, luminosity ratios and temperature
differences of the main-sequence visual binaries situated in the near Solar neighbourhood. An accurate
determination of the distributions of the binary properties is needed to provide observational
constraints to the various existing scenarios of binary star formation and offers a direct calibration
tool for  basic stellar properties. As the field main-sequence binaries serve as a reference for
various binary populations in different environments (e.g. in clusters, in metal-poor or star-forming
regions), it remains worthwhile to improve their statistics and the data obtained in the past  (at a
time when visual double stars were still "fashionable"), more particularly also for the  wide binaries 
which carry the largest angular momenta but which are also the ones most prone to dynamical disruption
(\cite{mat01}). For these purposes, monitoring of the (changing) astrometric parameters providing the
fundamental  binary data as well as precise measuring of the individual magnitudes and colour indices
of the components is absolutely needed. Accurate photometric differences allow the characterization of
the evolutionary stage of the components since basic properties such as luminosity/mass ratios as well
as temperature differences may be derived. 

The paper is structured in the following way: firstly, we describe the sample (Sect.~2), 
next we report on the observations, the reduction method and the astrometric calibration 
(Sect.~3). In Sect.~4 we present new astrometric and photometric measurements including 
a discussion of the errors. We compare our measurements to Hipparcos or older data, 
derive the properties of individual systems and discuss some unresolved systems in 
Sect.~5. A short summary of the observational results can be found in Sect.~6.

\section{Description of the sample}
We present high-accuracy relative astrometry and (B)VRI differential photometric data for a large
sample of visual double and multiple stars pertaining to the Hipparcos Catalogue (\cite{hc}). 
Complementary ground-based, multi-colour observations of the components of double stars observed 
during the Hipparcos mission are valuable because they provide accurate colour differences and 
independent monitoring of the relative position with the same high accuracy provided that the angular
separation is large enough. At the start of this project, a careful comparison between the CCDM
catalogue (\cite{ccdm}) on the one hand and the Hipparcos relative data of visual double stars 
(i.e. the Hipparcos Double and Multiple Systems Annex or Vol.~10, mainly Component Solutions (DMSA, 
Part C)) on the other hand showed that, in almost 10\% of the $\approx$ 9000 cases explored, a discrepancy was 
noted. This discrepancy can refer to the relative astrometry being discordant (with limits for target
selection as a function of angular separation as shown in Table~\ref{t1}) found in 3.2\% of the cases 
or to the relative photometry showing an excess of 1.2 mag at least in $\Delta$m (independent of the 
used filter) found in 6.4\% of the cases. This may seem a significant fraction whereas a comparison 
between Hipparcos and a sample of speckle binaries showed good agreement at the mas level (\cite{mi}). 
When the relative astrometry is obviously conflicting with the previous (generally much older) 
ground-based data for systems with larger angular separations, we can expect to be able to validate 
the space results and to show evidence of relative motion between the components, or a new component 
might have been detected by the satellite, or the space results are refuted permitting to correct 
the "ambiguous" solution proposed in the Hipparcos Catalogue (typically at an angular separation 
larger than 10\arcsec). When a large difference in differential magnitude is detected, one may also 
expect various reasons: a different signal in the considered passband, flux variability, a wrong 
component identification, a new component detection or an "ambiguous" Hipparcos solution. 
Further criteria adopted to select the programme double stars were:
\begin{itemize}
\item{accessibility from the Northern hemisphere ($\delta$ $>$ 0\degr)}
\item{they should be of 'intermediate' angular separation (1.5\arcsec~ $\le$ $\rho$ $<$ 15\arcsec)}
\end{itemize}
The initial target list consisted of 245 candidate visual double or multiple systems listed
in the Hipparcos Catalogue. Due to their apparent brightness, it appeared that several 
among these systems were also listed in the Catalogue of Nearby Stars (GJ, \cite{gj}).

\begin{table}
\caption{\label{t1} Adopted limits of target selection as a function of angular separation}
\begin{center}
\begin{tabular}{|c|@{}c@{}|@{}c@{}|@{}c@{}|@{}c@{}| }
\hline 
\mcol{1}{|c|}{Limit}
& \mcol{1}{c|}{$\rho \leq 1.5" $}
& \mcol{1}{c|}{$\rho \leq 5" $}
& \mcol{1}{c|}{$\rho \leq 12" $}
& \mcol{1}{c|}{$\rho > 12"$ \struut } \\ 
\hline
$\Delta \rho$/$\rho$ (\%) & 55 & 20 & 8 & 10 \struutup \\
Absolute  & 0.8\arcsec & 1.0\arcsec & 1.0\arcsec & $> 1.5$\arcsec \struutdown \\
\hline
\end{tabular} 
\end{center}
\end{table}   

From 2000 onward, we focused our attention specifically unto those visual double and multiple 
stars of our target list which belong the Catalogue of Nearby Stars (GJ, \cite{gj}) and/or have 
Hipparcos parallaxes larger than 0.04\arcsec. As before, we chose systems of `intermediate' 
angular separation fainter than apparent visual magnitude 7 and for which CCD observations 
can provide both accurate and complementary data on each component (cf. Paper~I). 
To complete our sample of nearby systems, we also added a small number of faint systems 
which are not included in the Hipparcos Catalogue (because they were beyond the Hipparcos 
magnitude limit; they will accordingly be designated by their Gliese number). 

The absolute parallax measurements obtained in space of the nearby systems from the sample 
are generally very accurate. In a few cases however, the ground-based parallaxes from 'The General 
Catalogue of Trigonometric Stellar Parallaxes' (\cite{van}, vALH) attain a higher accuracy and 
supersede the Hipparcos parallax. 

\section{Observations and data reduction}

\subsection{Instrumentation and limitations}                                         

The observations were performed at two observatories -- with the 2-m telescope of NAO (Rozhen, 
Bulgaria) and with the 0.6-m telescope of AOB (Belogradchik, Bulgaria). The main characteristics 
of the telescopes and their cameras are listed in Table~\ref{t2}. The telescope and the camera 
used at AOB are described in great detail in \cite{ba}. During this period, out of a total of 25 
allocated nights, some 35\% of the available time was effectively used. 

The CCD frames were taken from October 1998 to November 2004 through standard Johnson 
V and Cousins R, I filters. At a later stage at NAO, we also used the Johnson 
filter B. During the observations the exposure times were adjusted to get the highest 
possible counts for the primary (brighter) component; exposure times were usually several 
seconds long. On each night, a set of biases was taken every few hours. Flat-fields were 
obtained during evening and/or morning twilights: a set of 3 to 6 flats per filter was taken 
every night. Typical seeing values ranged from $1.5$ to $3$\arcsec.

\begin{table}[t]
\caption[]{\label{t2} Telescopes and instrumentation}
\begin{tabular}{rrr}
\noalign{\smallskip}
\hline
\noalign{\smallskip}
Site        & NAO Rozhen                    & AO Belogradchik               \\
\hline
Telescope   & 2-m Ritchy-Chr\'etien         &  60-cm Cassegrain             \\
CCD type    & Photometrics AT200            &  ST8                          \\
Chip        & Site SI003AB UV-AR            &  KAF 1600                     \\
Chip size   & $1024\times1024$ pixels       &  $765\times510$ pixels$^{1}$  \\
Pixel size  & $24\times24$ $\mu$m           &  $18\times18$ $\mu$m$^{1}$    \\
Scale       & 0.31\arcsec/pixel             &  0.49\arcsec/pixel$^{1}$      \\
Field       & $5.3\arcmin\times5.3\arcmin$  &  $6.2\arcmin\times4.2\arcmin$ \\
Gain        & 4.93 $e^-$/ADU                &  2.3 $e^-$/ADU                \\
RON         & 1.03 ADU/rms                  &  10 ADU/rms                   \\
\hline
\end{tabular}

$^{1}${With a binning factor of $2\times2$} \\
\end{table}                                                      
                                                             
\subsection{The reduction method}

All the primary reduction steps were performed using ESO-MIDAS standard routines. The frames 
were processed for bias and flat-field corrections. They include: subtraction of the residual 
bias pattern using a median master zero  exposure frame and flat-fielding using a median master 
flat-field frame. 

Next, we computed the angular separations (in pixel unit), the position angles and the magnitude
differences in the various filters for the components of the double stars. For this we used a
two-dimensional Moffat-Lorentz profile (\cite{mo}) fitting method. The code was developed within 
the ESO-MIDAS environment (\cite{cu}) and used in previous work (\cite{la01}, Pap.~I, Pap.~II). 

\subsection{The astrometric calibration}                                                 

To convert the instrumental angular separations and position angles into absolute values 
we applied the astrometric corrections as computed from stars observed in various standard 
astrometric fields (see Table~\ref{t3}). We measured the ($x, y$) positions and computed a 
multi-linear regression fit between the ($x, y$) positions and the catalogued ($\alpha, 
\delta$) values of the standard stars. Typically, we used 8-9 or more stars (cf. Col. 8 in 
Table~\ref{t3}) with reference coordinates in each field. On one occasion, only 3 stars were 
used. To compensate for this low number of reference stars (particularly in NGC~1647), we 
computed the mean calibration obtained from two such fields whenever possible.
We then determined the pixel scale and the orientation of the CCD chips (measured from North 
towards East). For this computation we made use of the software package Mira AP\footnote{The 
software Mira AP is produced by Axiom Research Inc., http://www.axres.com/} as well as of 
self-made codes. Both gave equivalent results for the same field. The adopted corrections are 
listed in Table~\ref{t3}. In column~9 we mention the source of the coordinates of the standard 
fields. The relative astrometric data were corrected using the appropriate values. The computed 
scale values of Table~\ref{t2} are not exactly the same as the nominal instrumental specifications 
(Sect.~3.1). 

\begin{table*}[t]
\caption[]{\label{t3} Astrometric calibration}
\begin{tabular}{lllllllll}
\hline
Observatory$^1$ & Astrometric    & Date & Scale          & $\sigma$(Scale) & Orient.   & $\sigma$(Orient.) &  N\_Stars & Source         \\
                & standard field &      & ($\arcsec$/px) & ($\arcsec$/px)  & ($\degr$) & ($\degr$)         &          & of coordinates \\
\noalign{\smallskip}
\hline
\noalign{\smallskip}
NAO   & M~15          &  1998 & 0.311 & 0.002 & -1.68   & 0.03   &  13       & Guide Star Catalogue (V7)   \\
NAO   & NGC~1647      &  2000 & 0.310 & 0.001 & -1.5    & 0.1    &  3$^2$     & \cite{ge}                   \\
NAO   & M~15          &  2001 & 0.310 & 0.001 & -2.04   & 0.02   &  8         & \cite{ca}                   \\
NAO   & M~67          &  2002 & 0.310 & 0.001 &  0.716  & 0.003  &  9         & Tycho Catalogue (\cite{hc}) \\
NAO   & M~67          &  2004 & 0.314 & 0.001 &  1.42   & 0.02   &  15+$^3$   & \cite{gi}                   \\
AOB   & NGC~1647      &  1998 & 0.493 & 0.01  &  2.29   & 0.1    &  10$^4$    & \cite{ge}                   \\
AOB   & M~15          &  1999 & 0.495 & 0.001 &  0.4    & 0.1    &  5         & Guide Star Catalogue (V7)   \\
AOB   & M~16          &  2000 & 0.493 & 0.001 & -1.62   & 0.04   &  9         & \cite{hi}                   \\
\hline
\end{tabular}

$(^1)$ NAO Rozhen: {2-m telescope}  --- AO Belogradchik: {0.6-m telescope}\\
$(^2)$ Field I  --- $(^3)$ Fields I and III  --- $(^4)$ Fields I and IV \\
\end{table*}

\section{CCD astrometry and photometry}
\subsection{Astrometric measurements}

The astrometric data are listed in Table~\ref{t4}. The first column mentions the Hipparcos 
identification number, followed by the component identification, the epoch (Bessel year), the 
number of frames, the angular separation ($\rho$) and the position angle ($\theta$) measured 
from North to East, with the respective standard errors $\sigma(\rho)$ and $\sigma(\theta)$. 
The values of $\rho$ and $\theta$ are the means of several frames measured in different filters. 
Typically, as many as 20-30 frames were obtained for each target, all filters included.

The standard errors of the mean values are quoted. These values are systematically better in 
the case of the NAO measurements: the mean uncertainties are 0.01$\arcsec$~in angular separation 
and 0.04$\degr$~in position angle. Such errors are typical for this range of angular separation 
(i.e. 'intermediate', \cite{la01}).
The resolution is also higher, as shown by the measurement of HIP~97237, with the smallest angular 
separation measured ($\rho$=1.4\arcsec). At AOB, the mean uncertainties are somewhat larger, i.e. 
0.04\arcsec~in angular separation and 0.26\degr~in position angle. Figure~1 illustrates 
the uncertainty in $\rho$ as a function of angular separation and of differential 
V magnitude between the components.

% Figure 1
\begin{figure}[t]
\label{fig1}
\resizebox{8cm}{!}{\includegraphics[angle=270]{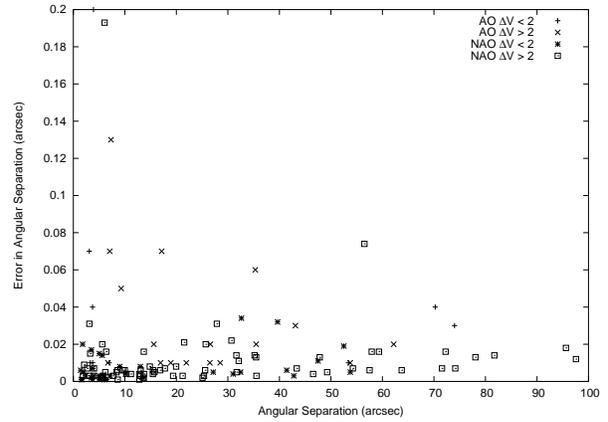}}           
\caption[]{Astrometric error vs. angular separation (in arcsec)}
\end{figure}

\subsection{Photometric data}

In Table~\ref{t5} we list the photometric magnitude and the colour differences in the following 
order: the Hipparcos number (col. 1), the component identification (col. 2), the heliocentric 
Julian Date (col. 3), the differential V magnitude ($\Delta$V) (col. 4) and the colour 
differences ($\Delta$B-$\Delta$V), ($\Delta$V-$\Delta$R) and ($\Delta$V-$\Delta$I) (cols. 5, 
7, 9, and 11), as well as the respective standard errors of the differences (cols. 6, 8, 10, and
12). As before, we consider that these values reflect the true colour differences between  
the components (e.g. $\Delta$(B-V)).

The standard errors of the mean values are quoted. The mean error of the differential V magnitude 
is 0.01~mag for the NAO observations while it is 0.03~mag for the AOB observations. Again, such 
mean internal errors are conform with expected values (\cite{la01}). Figure~2 illustrates 
the uncertainty in $\Delta$V as a function of angular separation and of the 
difference in V magnitude between the components. 

% Figure 2
\begin{figure}[t]
\label{fig2}
\resizebox{8cm}{!}{\includegraphics[angle=270]{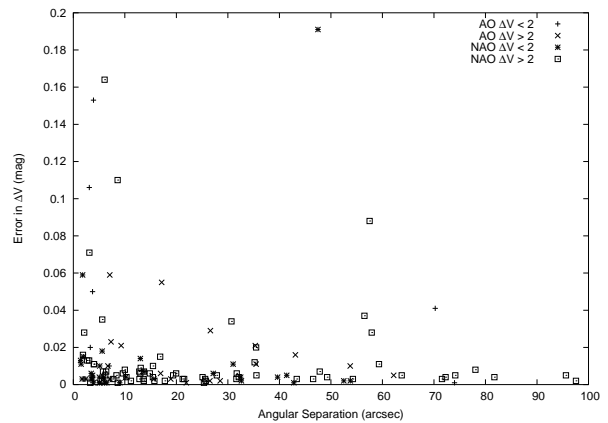}}             
\caption[]{Photometric error vs. angular separation (in arcsec)}
\end{figure}

\section{The nature of the association of individual systems and some unresolved systems}

\setcounter{table}{6}
\begin{table*}[t]
\caption[]{\label{t7} Orbital systems and comparison with the ephemeris based on the best fitting orbit(s)}
\begin{tabular}{lccccccccl}
\noalign{\smallskip}
\hline
\noalign{\smallskip}
Identifier & $\rho$$_{obs}$ & $\theta$$_{obs}$ & $\rho$$_{eph}$ & $\theta$$_{eph}$ &  $\delta$$\rho$ & $\delta$$\theta$ & 
$\Delta$Pos & Source \struutdown \\
\hline
HIP~473    & 6.079  & 178.49 &  6.068   & 182.61 &   0.011 &  -4.120 & 0.437 &  \cite{kiy01} \\ 
HIP~67422  & 3.30   & 170.70 &  3.320   & 173.48 &  -0.020 &  -2.780 & 0.162 &  \cite{he2}   \\ 
HIP~72659  & 6.72   & 314.20 &  6.636   & 317.28 &   0.084 &  -3.080 & 0.369 &  \cite{so99}  \\ 
HIP~79607  & 6.96   & 232.80 &  6.955   & 236.13 &   0.005 &  -3.330 & 0.404 &  \cite{ruy}   \\ 
HIP~88601  & 3.73   & 145.60 &  3.780   & 147.24 &  -0.050 &  -1.640 & 0.119 &  \cite{ruy}   \\ 
HIP~97237  & 1.418  &  46.98 &  0.940   &  82.54 &   0.478 & -35.560 & 0.852 & \cite{he4}    \\
HIP~97237  & 1.418  &  46.98 &  1.521   &  47.44 &  -0.103 &  -0.460 & 0.104 & \cite{so99}   \\
\hline
\end{tabular}
\end{table*}

In Table~\ref{t6} we show the difference in relative position, $\Delta$Pos, (col.~9) between the 
new values and those from the Hipparcos Catalogue (for mean epoch 1991.25). When no Hipparcos data 
were available, we used the (sometimes much) older data from the CCDM (\cite{ccdm}). Also listed are the 
catalogue's epoch (col.~4), the published position angle (col.~5) and angular separation (col.~6), 
the computed differences in $\rho$ (col.~7) and $\theta$ (col.~8). Lastly, we shortly comment the 
nature of the association of the system using a number of codes ('S'=stable; 'L'=showing a linear 
relative motion (optical system);'M'=showing (orbital) motion; and 'O'=with known orbital motion).
Code 'L' is used when the difference in relative position is compatible with the (measured or
estimated) relative proper motion of the components considered while code 'M' is assigned when this 
is not the case. The derived properties for some of the orbital binaries in the sample under study 
have already been published (Pap.~II). Therefore, we will not include these binaries in the discussion, 
unless they are part of a complex (e.g. multiple) system. However, they are included in Table~\ref{t7} 
where a comparison is made with the ephemeris computation based on previously known orbits (\cite{wds1}).

The comments addressing specific systems of Tables~\ref{t6} and~\ref{t7}, which for various reasons 
do not form stable configurations (i.e. are not fixed systems) can be found in the Appendix.

In Tables~\ref{t8} and \ref{t9}, we list newly derived binary properties for 20 bound systems of Table~\ref{t6}. 
We provide the system's (B-V) colour index (col. 2), the component colours (cols. 3 and 4), the bolometric 
correction difference (col. 5), the bolometric magnitude difference (col. 6) and the subsequently derived
fractional mass, $\beta$, (col. 7) (Table~\ref{t8}). In Table~\ref{t9}, we provide (lower limits of) the linear 
separations based on the observed angular separations (generally corresponding to larger semi-major axes) 
by making use of the most precise parallaxes known to-date (a note in col.~13 indicates the use of a ground-based 
trigonometric parallax).

\begin{table*}[t]
\caption[]{\label{t8}  Basic binary properties: component colours and masses }
\begin{tabular}{rcccccccccc}
\noalign{\smallskip}
\hline
\noalign{\smallskip}
Identifier & (B-V)$_{AB}$ & (B-V)$_{A}$ & (B-V)$_{B}$ & $\Delta$BC & $\Delta$M$_{Bol}$ & $\beta$ & M$_{A+B}$     & $\sigma($M$_{A+B})$ & M$_A$  & M$_B$ \\
           &              &             &             &            &                   &         & $(M_{\odot})$ & $(M_{\odot})$       & $(M_{\odot})$ & $(M_{\odot})$ \\
\hline
HIP~473$^{1}$    & 1.443 & 1.444 & 1.442 &  0.004 &  0.064 & 0.496 & 1.496 & 0.27  & 0.75 & 0.74 \\
HIP~1860         & 1.450 & 1.453 & 1.402 &  0.109 &  2.973 & 0.328 & --    & --    & --   & --   \\
HIP~15844        & 1.500 & 1.512 & 1.463 &  0.130 &  1.407 & 0.416 & --    & --    & --   & --   \\
HIP~17666        & 0.799 & 0.751 & 0.887 & -0.113 &  0.474 & 0.471 & --    & --    & --   & --   \\
HIP~29316        & 1.450 & 1.446 & 1.469 & -0.053 &  1.575 & 0.406 & --    & --    & --   & --   \\
HIP~43422        & 1.355 & 1.216 & 1.521 & -0.578 & -0.536 & 0.532 & --    & --    & --   & --   \\
HIP~44295        & 1.285 & 1.191 & 1.395 & -0.308 & -0.244 & 0.515 & --    & --    & --   & --   \\
HIP~92836        & 1.370 & 1.367 & 1.429 & -0.118 &  3.186 & 0.317 & --    & --    & --   & --   \\
HIP~97237        & 1.720 & 1.739 & 1.709 & -0.612 & -0.003 & 0.500 & 0.463 & 0.086 & 0.23 & 0.23 \\ 
HIP~110640$^{1}$ & 1.190 & 1.162 & 1.480 & -0.530 &  1.848 & 0.390 & --    & --    & --   & --   \\
HD~23713         & 0.544 & 0.156 & 1.346 & -0.849 & -0.638 & 0.538 & --    & --    & --   & --   \\
\hline
\end{tabular}

$^{(1)}${cf. also Pap.~II}
\end{table*}

\begin{table*}[t]
\caption[]{\label{t9} Basic binary properties: (lower) linear separations for 20 bound systems}
\begin{tabular}{rcccrrrrrrrrc}
\noalign{\smallskip}
\hline
\noalign{\smallskip}
\mcol{1}{c}{Identifier}& \mcol{1}{c}{Cp}& \mcol{1}{c}{Epoch}& \mcol{1}{c}{N\_ima}& \mcol{1}{c}{$\rho$}&\mcol{1}{c}{$\sigma_{\rho}$}& \mcol{1}{c}{$\theta$}& \mcol{1}{c}{$\sigma_{\theta}$}& \mcol{1}{c}{$\pi$}& \mcol{1}{c}{$\sigma_{\pi}$}& 
\mcol{1}{c}{A$_{low}$}& \mcol{1}{c}{$\sigma_{A}$} & \mcol{1}{c}{Note}\\
\mcol{1}{c}{}& \mcol{1}{c}{}& \mcol{1}{c}{($Bessel~yr$)}& \mcol{1}{c}{}& \mcol{1}{c}{($\arcsec$)}&\mcol{1}{c}{($mas$)}& \mcol{1}{c}{($\degr$)}& \mcol{1}{c}{($\degr$)}& \mcol{1}{c}{($mas$)}& \mcol{1}{c}{($mas$)}& 
\mcol{1}{c}{($A.U.$)}& \mcol{1}{c}{($A.U.$)} & \mcol{1}{c}{}\\
\hline 
 HIP~473    & B & 2001.8583 & 24 &  6.079 &  1 & 178.49 & 0.01 &  85.10 & 2.74  &  71   & 2    &         \\
 HIP~1860   & B & 2001.8611 & 58 & 11.178 &  4 &  58.13 & 0.03 &  50.71 & 2.72  & 220   & 12   &  vALH   \\
 HIP~4258   & B & 2001.8584 & 24 &  6.488 &  1 &  66.52 & 0.01 &   8.59 & 2.24  & 755   & 202  &  vALH   \\
 HIP~9275   & B & 2000.8238 & 40 &  3.926 &  3 &  54.57 & 0.07 &  33.53 & 5.29  & 117   & 18   &         \\
 HIP~15844  & B & 2000.8213 & 23 &  2.397 &  3 & 340.90 & 0.05 &  50.54 & 4.66  &  47   & 4    &         \\
 HIP~17666  & B & 2000.8267 & 48 &  7.139 &  3 &  51.81 & 0.02 &  40.83 & 2.24  & 175   & 10   &         \\
 HIP~21088  & B & 2000.8213 & 15 &  9.022 &  8 &  61.81 & 0.08 & 180.60 & 0.80  &  49.9 & 0.3  &  vALH   \\
 HIP~22715  & B & 2000.8324 &  5 &  3.980 &  3 & 216.32 & 0.03 &  37.09 & 1.37  & 107   & 4    &         \\
 HIP~29316  & B & 2000.8216 & 20 &  1.803 & 20 &  27.04 & 0.50 &  97.90 & 3.90  &  18   & 1    &  vALH   \\
 HIP~39896* & B & 2004.8843 & 15 & 13.750 &  2 & 240.32 & 0.02 &  48.26 & 3.16  & 285   & 19   &         \\
 HIP~41824  & B & 2000.8353 & 18 & 10.149 &  4 & 344.76 & 0.01 &  78.05 & 5.69  & 130   & 10   &         \\
 HIP~43422  & B & 2002.9106 & 12 &  1.717 &  1 & 153.07 & 0.04 &  31.24 & 19.30 &  55   & 34   &         \\
 HIP~44295* & B & 2004.8845 & 15 &  5.100 &  3 & 180.80 & 0.01 &  54.57 & 3.21  &  93   & 5    &         \\
 HIP~92836  & B & 2001.8637 & 48 &  4.021 &  7 &  32.92 & 0.02 &  50.30 & 2.70  &  80   & 4    &         \\
 HIP~97237  & B & 2000.8342 & 19 &  1.418 &  6 &  46.98 & 0.21 &  94.70 & 4.40  &  15   & 1    &  vALH   \\
 HIP~110640 & B & 2001.8582 & 55 &  2.103 &  9 & 220.94 & 0.27 &  46.74 & 1.66  &  45   & 2    &         \\
 HIP~113437 & B & 2001.8610 & 24 &  1.515 &  1 & 252.96 & 0.09 &   8.19 & 1.52  & 185   & 34   &         \\
 HD~23713   & B & 2000.8296 & 15 &  1.935 &  5 & 126.96 & 0.11 &  45.00 & 15.00 &  43   & 14   &  vALH   \\
 GJ~1047    & C & 2000.8321 & 24 & 31.025 &  4 & 233.25 & 0.04 &  46.20 & 3.60  & 672   & 52   &  vALH   \\
 GJ~1245    & B & 2000.8344 & 20 &  7.035 &  3 &  79.63 & 0.02 & 220.20 & 1.00  &  32.0 & 0.2  &  vALH   \\
\hline  
\end{tabular}

$^*$ This flag denotes another epoch for the same target (cf. Table 4)
\end{table*}

In seven cases, the companion (component~B) was not detected and thus not measured. This was the case
for the following systems: HIP~8414 (1991: $\rho$=1.7\arcsec), HIP~21765 (1960: $\rho$=2\arcsec), 
HIP~30920, HIP~101150, HIP~105747 (1991: $\rho$=0.1\arcsec), GJ~1047~AB (1966: $\rho$=1\arcsec) and 
GJ~1103~AB. In six cases (not GJ~1103), the angular separation previously measured was (well) below 2\arcsec. 
We also did not resolve the binary GJ~1103~AB which was measured with an angular separation of about 3\arcsec~in 
1960. On the other hand, we know that the adopted lower limit of 1.5\arcsec~in angular separation can be 
reached under good circumstances (depending e.g. on seeing and on the observed magnitude difference, e.g. 
HIP~97237 and HIP~113437). We therefore claim that the most probable reason for non-detection in these 
systems is the fact that the binaries presently have an angular separation equal to or below 1.7\arcsec~ 
(1.5\arcsec~for $\Delta$m $<$ 1 mag). They were presently unresolved by the adopted CCD technique. In the 
case of HIP~30920, component B is actually situated at an angular separation of 1.3\arcsec~(\cite{wds1}), 
whereas the Hipparcos angular separation of HIP~101150 was only 0.8\arcsec~(epoch 1991.25). 

HIP~12781 and HIP~28368 are two newly discovered doubles according to the Hipparcos Catalogue (\cite{hc})
for which we report no detection of an additional component with a separation above 1.7\arcsec~ 
(1.5\arcsec~for $\Delta$m $<$ 1 mag).

\section{Conclusions}

We provided high-accuracy astrometric and photometric measurements for 71 visual systems, 
of which 27 are nearby (with parallax less than 30 pc) and 30 are multiple systems. In three 
additional cases, the binary systems remained unresolved. 

From a comparison with the relative positions from the Hipparcos Catalogue (for mean epoch 1991.25) 
or, when no Hipparcos data were available, with the data from the CCDM Catalogue, we evaluated the 
physical status of 55 systems. To summarize, we found that:    \\
 - 8 systems show a linear relative motion (optical),          \\
 - 17 systems are true binaries showing motion,                \\
 - 22 systems show a fixed configuration,                      \\
 - 5 systems probably show a fixed configuration,              \\
 - 3 binaries have a published orbit.                                      

Comparison of the new measurements and the ephemerides computed with the orbits found in the 
literature shows a reasonably good agreement in the case of three binaries (HIP~67422, HIP~88601, and 
HIP~97237) but too large residuals in the other cases (HIP~473, HIP~72659 and HIP~79607)(cf. 
Table~\ref{t7}). The long-term monitoring of these orbital pairs should be pursued with an adequate 
angular resolution. 

Two new visual double stars of intermediate separation were also discovered, one of which 
(HD~218587) was already reported elsewhere (\cite{str}). 23 new components of known systems were 
measured and basic binary properties were newly determined for 20 bound systems. 

In the case of two orbital binaries, we derived a full set of fundamental parameters including 
new component colours and corresponding component masses.

\begin{acknowledgements}
PL and AS acknowledge financial support from the Belgian Science Policy and from the Bulgarian 
Academy of Sciences through bilateral project "Astrometric, spectroscopic and photometric follow-up 
of binary systems" (reference BL/33/B11). PL is grateful to Prof. K. Panov for the allocation of
telescope time and for the help provided by the operators at the NAO, Rozhen. The CCD ST-8 at the 
AO Belogradchik was financed by the A. von Humboldt foundation (Germany). We further thank the 
anonymous referee for a careful reading and many useful suggestions. This research made 
extensive use of various data bases among which SIMBAD and VIZIER, operated at the CDS (Strasbourg, 
France) as well as of the catalogues maintained at the U.S. Naval Observatory 
[http://ad.usno.navy.mil/wds/dsl.html] (Washington, DC) and the Guide Star Catalogue 
[http://www-gsss.stsci.edu/gsc/GSChome.htm]

\end{acknowledgements}

{}

\newpage

ON-LINE DATA

\break
Appendix A: List of comments addressing specific systems of Tables~\ref{t6} and~\ref{t7} which
for various reasons do not show a fixed binary configuration \\

{\em HIP~473:} GJ~4~AB is a nearby orbital pair ($\pi_{Hip}$ = 85.10 $\pm$ 2.74 mas) in a multiple system 
showing a high common proper motion (comp A: total proper motion (pm) of 0.891$\arcsec$/yr in the direction 
100$\degr$, comp B:  total pm of 0.853$\arcsec$/yr in the direction 101$\degr$), also for example HIP~428 
(comp F: total pm of 0.883$\arcsec$/yr in the direction 100$\degr$) (\cite{hc}). The relative position of 
component E is consistent with a change in the position of GJ~4~AB over a period of 77 years.  
The comparison with the ephemeris of \cite{kiy01} is less concordant than previously (in Pap. II we already
stated that this orbit may not be definitive). The observed change of rate of the position angle is opposite 
to the expected one. The system was recently also measured by \cite{pav05} who have good agreement with 
the proposed ephemeris but less accuracy than our data. For this reason also, we thoroughly checked 
the determination of the zero point of the orientation angle.

{\em HIP~1397:} shows a significant proper motion, definitely L-type. This is confirmed by the Hipparcos
difference in proper motion of the order of 0.1"/yr between the "components" in the direction 10$\degr$ 
which is fully compatible with the observed change in relative position of almost 1" (i.e. 
$\Delta$Pos=0.942") over slightly more than 10 years ($\pi_{Hip}$ = 11.30 $\pm$ 1.39 mas, \cite{hc}).

{\em HIP~1860:} GJ~1010~AB, a nearby system with a high proper motion, possibly M-type (total pm of 0.800$\arcsec$/yr
in the direction 273$\degr$ and $\pi_{tr}$ = 62.8 $\pm$ 4.0, \cite{van}). There is no double-star solution mentioned 
in the Hipparcos Catalogue (since it is listed in DMSA/X, Part X which contains the stochastic solutions for 
objects for which no single- nor double-star solution could be found in reasonable agreement with the standard 
errors of the Hipparcos observations)(\cite{hc}). It forms a common proper-motion binary listed in (\cite{go}) 
(=NLTT 1186 and 1189).

{\em HIP~3589:} shows a significant proper motion, most probably L-type. This is confirmed by the Hipparcos
difference in proper motion of the order of 0.1"/yr between the "components" ($\pi_{Hip}$ = 20.56 $\pm$ 1.69 mas).

{\em HIP~4258:} GJ~1023~AB shows a significant proper motion (total pm of 0.100$\arcsec$/yr in the direction 
250$\degr$, \cite{hc}), most probably M-type. It has $\pi_{tr}$ = 53.0 $\pm$ 15.9 mas (\cite{van}) whereas 
$\pi_{Hip}$ = 8.59  $\pm$ 2.24 mas. 
% 0.094  254    0.0530  15.9

{\em HIP~7495:} the new measurement confirms rather well the Hipparcos "alternative" solution (\cite{hc}).

{\em HIP~9275:} GJ~1041~AB has a high proper motion (total pm of 0.260$\arcsec$/yr in the direction 
84$\degr$, \cite{hc}), probably M-type ($\pi_{Hip}$ = 33.53 $\pm$ 5.29 mas). There is no differential
proper motion known. 

{\em HIP~9867:} GJ~84.2~AB, also BD +44$\degr$423 (not BD +44$\degr$422), is a high proper motion (pm of 
0.510$\arcsec$/yr in the direction 148$\degr$ (\cite{van})) and a possible EA variable star (=V 374 And) 
($\pi_{tr}$ = 53.5 $\pm$ 5.2 mas). Component B is in the field but was at first not identified (this component 
is mentioned in the CCDM but not in the NLTT catalogue). GJ 84.2 AB (Wor 1) is evidently an optical pair. 
Though they are background stars, two "components" (B?,E') were also measured. We think that background star
B? probably corresponds to component B (located at (307$\degr$,4.4$\arcsec$) in 1959). Note that another recent 
position of component B? has also been attributed to component B (\cite{wds1}). The relative position is concordant 
with a change in the position of component A over a period of 42 years. The relative position of component 
C is also concordant. The Hipparcos stochastic solution was rejected because it had a "cosmic error" greater 
than 100 mas (\cite{hc}). 

{\em Unresolved system HIP~12781:} GJ 109 (=LHS 1439), also the flare star VX Ari, has a huge proper motion (total pm of 
0.924\arcsec/yr in the direction 114\degr (\cite{van}) and $\pi_{Hip}$ = 127.3 $\pm$ 4.2 mas). It is 
included in the Double and Multiple Systems Annex with a variable component (DMSA/V, Part V which contains
the VIM solutions for objects where the duplicity has been inferred by a photocentric motion caused by the 
variability of one of the components (i.e. Variability Induced Movers))(\cite{hc}). We report no detection of 
an additional component with a separation above 1.7\arcsec~ (1.5\arcsec~ for $\Delta$m $<$ 1 mag). 

{\em HIP~15844:} GJ~140~AB has a high proper motion (pm of 0.253$\arcsec$/yr in the direction 121$\degr$, \cite{hc}), 
the binary motion is confirmed ($\pi_{Hip}$ = 50.54 $\pm$ 4.66 mas) (Pap.~II). Component C was not in the field
(=NLTT 10808 at (118.2$\degr$,99.5$\arcsec$)) but forms a common proper-motion pair with component A (\cite{go}). 

{\em HIP~17102:} Wo~9119~AB. This measurement does not agree with the Hipparcos nor with the Hipparcos "alternative" 
solution. The observation is nevertheless consistent with it as the difference with the Hipparcos solution is almost 
exactly 1 gridstep (equal to 1.2$\arcsec$)($\pi_{Hip}$ = 20.03 $\pm$ 2.14 mas).

{\em HIP~17666:} GJ~1064~AB, also a variable star with a huge proper motion (comp A: pm of 1.377$\arcsec$/yr in the 
direction 154$\degr$, comp B: pm of 1.384$\arcsec$/yr in the direction 155$\degr$, \cite{hc}), shows a binary motion 
($\pi_{Hip}$ = 40.83 $\pm$ 2.24 mas). There is an important difference in position angle between the two observations 
(this work; Pap. II). A note in the Hipparcos Catalogue reports "Possibly E type. The double-star analysis indicates 
that it is probably the fainter (B) component which is variable."   

{\em HIP~21088:} GJ~169.1~AB has a huge proper motion (pm of 2.383$\arcsec$/yr in the direction 145$\degr$; 
$\pi_{tr}$ = 180.6 $\pm$ 0.8 mas, \cite{van}). Both measurements (this work; Pap~II) agree very well. Compared to 
the Hipparcos double-star solution, the system shows a distinct orbital motion (M-type) in agreement with older 
data. It also has an "alternative" Hipparcos solution which is less consistent. Since $\Delta$Pos is however close 
to 1 gridstep, the Hipparcos double-star solution should be treated with caution. Both components (=NLTT 13373 and 
13375) form a common proper-motion binary (\cite{luy79}).

{\em HIP~22715:} GJ~2035~AB has a high proper motion (pm of 0.197$\arcsec$/yr in the direction 131$\degr$,
\cite{hc}). The Hipparcos Catalogue gives no double-star solution ($\pi_{Hip}$ = 37.09 $\pm$ 1.37 mas). 
The proper motion applied over a period of 100 years does not explain the relatively small value of 
$\Delta$Pos of 1.4" (we would expect a 10-fold increase if it were an optical binary and the 
change in relative position was entirely caused by the difference in proper motion between the companions). 
Orbital motion is possibly detected with a rate in position angle of about 0.2\degr/yr.

{\em Unresolved system HIP~28368:} NN 3371 A has a high proper motion (pm of 0.253$\arcsec$/yr in the direction 
177$\degr$ (\cite{hc}). The star was first treated as a double in the Hipparcos Catalogue (DMSA/X) but later on reprocessed 
as a single star ($\pi_{Hip}$ = 74.17 $\pm$ 1.82 mas). No new component at a separation above 1.7\arcsec~ was found in the 
vicinity. Component B, situated at (119.6\degr,161.2\arcsec), was not in the field of view centered on the primary. 

{\em HIP~29316:} GJ~228~AB has a huge proper motion (pm of 0.970$\arcsec$/yr in the direction 176$\degr$ 
and $\pi_{tr}$ = 97.9 $\pm$ 3.9 mas, \cite{van}), definitely M-type (orbital) motion. There is a notable 
difference in angular separation with the Hipparcos double-star solution.
{\em HIP~31635:} GJ~239~A has a huge proper motion (pm of 0.844$\arcsec$/yr in the direction 293$\degr$, 
\cite{hc}), definitely with L-type motion due to the large differential proper motion between the "components". 
It furthermore has no Hipparcos double-star solution ($\pi_{Hip}$ =  101.59 $\pm$ 2.35 mas). 

{\em HIP~34222:} GJ~265~A has a significant proper motion (pm of 0.122$\arcsec$/yr in the direction 205$\degr$, 
\cite{van}), definitely L-type motion based on our measurements obtained at two different epochs. It has no 
Hipparcos double-star solution ($\pi_{Hip}$ =  41.63 $\pm$ 2.16 mas).

{\em HIP~39721:} Wo~9251~AB has a significant proper motion (pm of 0.136$\arcsec$/yr in the direction 181$\degr$,
\cite{van}), possibly M-type ($\pi_{Hip}$ = 24.85 $\pm$ 3.92 mas). There is a notable difference in angular 
separation with the Hipparcos double-star solution.

{\em HIP~39896:} GJ~1108~AB (pm of 0.196$\arcsec$/yr in the direction 195$\degr$ (\cite{van})) has a wrong Hipparcos 
double-star solution ($\pi_{Hip}$ = 48.26 $\pm$ 3.16 mas). Our two measurements are however consistent with the older 
CCDM data at a separation of 13$\arcsec$. It is also the variable star FP~Cnc.

{\em HIP~41824:} GJ~2069~AB has a significant proper motion (pm of 0.246$\arcsec$/yr in the direction 249$\degr$ and 
$\pi_{Hip}$ = 78.05 $\pm$ 5.69 mas, \cite{hc}). It has no Hipparcos double-star solution (but is in DMSA/V). Component 
A is the variable star CU~Cnc. Compared to the older data, the configuration is almost similar. Both components 
(=NLTT 19685 and 19684) form a common proper-motion binary (\cite{luy79}). This system is actually quintuple with three 
recently resolved new close components (\cite{beu}).

{\em HIP~43422:} GL~323~AB has a significant proper motion (pm of 0.146$\arcsec$/yr in the direction 271$\degr$ and 
$\pi_{Hip}$ = 31.24 $\pm$ 19.30 mas, \cite{hc}). It has a Hipparcos stochastic solution only (DMSA/X). It may present
orbital motion.

{\em HIP~44295:} GJ~1120~AB is a nearby system with a high proper motion (pm of 0.343$\arcsec$/yr in the direction 
201$\degr$ (\cite{van}) and $\pi_{Hip}$ = 54.57 $\pm$ 3.21 mas). A clear binary motion was detected. From a comparison
of our two measurements and the Hipparcos double-star solution, we obtain a decrease of 0.2$\degr$/yr in position 
angle, fully consistent with the rate of change detected by Hipparcos. 

{\em HIP~92836:} GJ~734~AB is a nearby system with a significant proper motion (pm of 0.120$\arcsec$/yr in the direction 
95$\degr$ and $\pi_{tr}$ = 61.5 $\pm$ 7.6 mas, \cite{van}). Orbital motion is possibly detected. One component is variable 
(V1436~Aql). The Hipparcos stochastic solution was rejected because it had a cosmic error greater than 100 mas (\cite{hc}). 
It was reprocessed as a single star later on (pm of 0.120$\arcsec$/yr in the direction 9$\degr$ and $\pi_{Hip}$ = 50.30 
$\pm$ 2.70 mas).

{\em HIP~95071:} GL~754.1~BA is a nearby system with a significant proper motion (pm of 0.199$\arcsec$/yr 
in the direction 198$\degr$ and $\pi_{tr}$ = 99.2 $\pm$ 2.5 mas, \cite{van}). Component B (= NLTT 47693) is 
variable (NSV 11920). It has a Hipparcos stochastic solution only (DMSA/X) ($\pi_{Hip}$ = 89.08 $\pm$ 7.16 mas). 
Orbital motion is clearly detected between components B and Q. There is a fainter common proper-motion companion 
showing an almost fixed configuration over more than a century (comp A = NLTT 47691) (\cite{go}).  

{\em HIP~97237:} GL~766~AB has a huge proper motion (total pm of 1.226$\arcsec$/yr in the direction 181$\degr$ 
and $\pi_{tr}$ = 94.7 $\pm$ 4.4 mas (\cite{van})). The orbital motion with respect to 40 years ago is clearly detected 
with a rate of change of 0.5$\degr$/yr in position angle. The Hipparcos stochastic solution was rejected because it 
had a cosmic error greater than 100 mas (\cite{hc}). Two orbits exist for this binary (Kui 95) in the literature. 
From the comparison in Table~\ref{t7}, only the orbit by \cite{so99} (P$_{orb}$= 228 yr) is reliable 
($\Delta$Pos = 0.1\arcsec). 
 
{\em HIP~101150:} Wo~9697~AB is a nearby system with a significant proper motion (pm of 0.195$\arcsec$/yr 
in the direction 228$\degr$ (\cite{van}) and $\pi_{Hip}$ = 43.24 $\pm$ 4.37 mas). We were not able to resolve 
the orbiting companion at a separation below 1$\arcsec$ (\cite{hc}). The position of the known component C has 
shifted due to the high differential proper motion over the last 40 years (see also the different parallax 
attributed by \cite{van}). Another "component" (D') was measured, but it has no link with the known component 
D. This binary needs further monitoring using the speckle-interferometric technique.

{\em HIP~104210:} shows a distinct motion of L-type. 
Previous analysis of all available data (\cite{wds1}) suggested that this is an optical pair. Using the colour 
difference between the components, we concluded that the secondary is a foreground star probably lying within 
100 pc (\cite{str}). The difference in relative position of about 0.7\arcsec with the Hipparcos value is entirely 
caused by the differential proper motion (0.089$\arcsec$/yr in the direction 40$\degr$) over the interval of 
7.5 years (\cite{hc}). 

{\em HIP~105421:} is a fixed system. The errors are much smaller than those of the Hipparcos Catalogue.

{\em HIP~108888:} is most probably a fixed system. Component A is the variable star V394 Lac. This measurement 
does not agree with the Hipparcos double-star solution. However, the difference in relative position is very
probably due to a gridstep ambiguity affecting the determination of $\rho_{Hip}$.

{\em HIP~108892:} $\Delta\rho$ is a bit large for a fixed system (though only at the 2$\sigma_{Hip}$-level). 
Our two observations do concord in angular separation, but not in position angle. One component is the pulsating
variable star V378~Lac. In view of the overall consistency with the old CCDM relative position however, it
probably is a fixed system.

{\em HIP~110326:} shows a distinct motion of L-type. There is a difference in relative position of 
about 0.4\arcsec~ with the Hipparcos value (\cite{hc}). It is most probably an optical pair whose 
components are moving apart at the relative speed of about 0.06-0.07\arcsec/yr.

{\em HIP~110640:} GL~857.1~AB has a significant proper motion (pm of 0.200$\arcsec$/yr in the direction 
244$\degr$ (\cite{van}) and $\pi_{Hip}$ = 46.74 $\pm$ 1.66 mas). The orbital motion was described in 
Pap~II. The position of component C has shifted in agreement with the high differential proper motion over 
more than one century. Our differential photometric data however indicates that "component" C is about 1 mag 
fainter than expected. 

{\em HIP~111279:} is probably a fixed system. Component A is HIP~111277 for which the Hipparcos stochastic 
solution was rejected because it had a cosmic error greater than 100 mas (\cite{hc}). The difference 
with the double-star solution proposed by Hipparcos (DMSA/C) approximates 1 gridstep. Our measurements 
agree very well with the older CCDM data. 

{\em HIP~113411:} is perhaps a fixed system. Our observation does not agree with the Hipparcos solution
even though this target has insignificant proper motion. The accuracy of our measurement is not very high, 
therefore an extra observation is needed to confirm whether some kind of relative motion is present or not.

{\em HIP~113437:} has a significant proper motion (pm of 0.096$\arcsec$/yr in the direction 74$\degr$ and 
$\pi_{Hip}$ = 8.19 $\pm$ 1.52 mas). Orbital motion has probably been detected since $\Delta \theta$ is 
important.

{\em HD~23713:} Cou~80~AB has an intermediate proper motion (pm of 0.041$\arcsec$/yr in the direction 
166$\degr$ and $\pi_{tr}$ = 45.0 $\pm$ 15.0 mas, \cite{van}). The comparison with older CCDM data indicates
orbital motion.

{\em GJ~1047~AB:} (=NLTT 7710) has a huge proper motion (pm of 0.919$\arcsec$/yr in the direction 128$\degr$ 
and $\pi_{tr}$ = 46.2 $\pm$ 3.6 mas). Component C (=NLTT 7708) shares the same proper motion (\cite{van}).

{\em GJ~1103~AB:} has a high proper motion (pm of 0.766$\arcsec$/yr in the direction 161$\degr$ and 
$\pi_{tr}$ = 114.0 $\pm$ 3.3 mas, \cite{van}). Component B (=NLTT 18546) formerly situated at (78.0\degr,
3.0\arcsec) forms a common proper-motion pair with component A (=NLTT 18545) (\cite{luy79}) but was 
not detected (with $\Delta$m$_{R}$ $\approx$ 2.5 mag). The other "components" (C',D',E') have no physical 
link with the binary system. 

{\em GJ~1245~AB:} has a high proper motion (pm of 0.731$\arcsec$/yr in the direction 143$\degr$ 
and $\pi_{tr}$ = 220.2 $\pm$ 1.0 mas, \cite{van}) and shows a clear orbital motion. Component B (=NLTT 48414) 
forms a common proper-motion pair with component A (=NLTT 48415) (\cite{luy79}) and shows a slight
change in position angle since 1997. There is a low mass companion (estimated to 0.1M$_{\odot}$) close to 
GJ~1245~A (\cite{gj}). 

\setcounter{table}{3}
\begin{table*}[t]
\caption[]{\label{t4} Astrometry of the observed stars}
\begin{tabular}{llcccccccl}
\noalign{\smallskip}
\hline
\noalign{\smallskip}
Identifier & Cp  & Epoch             & N\_frames & $\rho$     & $\sigma$($\rho$) & $\theta$   & $\sigma$($\theta$) & Telesc.$^1$ & Remark/ \\
        &       & ($Bessel~yr$) &    &($\arcsec$) &($\arcsec$)       & ($\degr$ ) & ($\degr$)          &             & Other identifier    \\
\hline 
HIP~473     & B & 2001.8583 &  24    &  6.079 &   0.001 &  178.49 &  0.01 &  2    & GJ 4 AB              \\
HIP~473     & C'& 2001.8583 &  23    &  9.628 &   0.006 &   11.66 &  0.02 &  2    & Not a true cmp       \\
HIP~473     & E & 2001.8583 &  23    & 54.265 &   0.007 &  352.04 &  0.01 &  2    &                      \\
HIP~1397    & B & 1998.7948 &  26    &  8.368 &   0.005 &  221.67 &  0.03 &  2    &                      \\ 
HIP~1860    & B & 2001.8611 &  58    & 11.178 &   0.004 &   58.13 &  0.03 &  2    & GJ 1010 AB  (DMSA/X) \\
HIP~3589    & B & 1999.7727 &  12    &  3.10  &   0.07  &   83.3  &  1.4  &  0.6  &                      \\ 
HIP~3589    & C & 1999.7727 &  13    & 43.10  &   0.03  &   94.2  &  0.1  &  0.6  &                      \\
HIP~3589    & D'& 1999.7727 &  13    & 17.15  &   0.07  &   70.8  &  0.1  &  0.6  & Not a true cmp       \\
HIP~4258    & B & 2001.8584 &  24    &  6.488 &   0.001 &   66.52 &  0.01 &  2    & GJ 1023 AB           \\ 
HIP~7495    & B & 2001.8638 &  55    &  1.859 &   0.003 &  303.84 &  0.10 &  2    &                      \\
HIP~7495    & C & 2001.8638 &  55    & 13.645 &   0.002 &  300.14 &  0.01 &  2    & New cmp              \\
HIP~8414    & B & 2001.8642 &   0    &  -     &   -     &    -    &   -   &  2    & B not detected       \\ 
HIP~8414    & C & 2001.8642 &   2    & 53.172 &   0.024 &  322.72 &  0.01 &  2    &                      \\ 
HIP~9275    & B & 2000.8238 &  40    &  3.926 &   0.003 &   54.57 &  0.07 &  2    & GJ 1041 AB           \\ 
HIP~9488    & B & 2001.8642 &  24    & 17.767 &   0.007 &   11.15 &  0.01 &  2    & Wo 9067 AB           \\ 
HIP~9867    & B?& 2000.8267 &  52    & 25.351 &   0.003 &  323.51 &  0.01 &  2    & probably B, previously $\rho$=4.4"\\
HIP~9867    & C & 2000.8267 &  52    &  9.996 &   0.006 &  230.89 &  0.02 &  2    & A=GJ 84.2            \\
HIP~9867    & E'& 2000.8267 &  52    & 42.804 &   0.003 &  287.73 &  0.01 &  2    & Not a true cmp       \\
HIP~10023   & B & 1998.7948 &   5    &  6.199 &   0.005 &  254.53 &  0.01 &  2    &                      \\
HIP~11390   & B & 1999.7728 &  13    & 74.00  &   0.03  &   31.7  &  0.1  &  0.6  & A=VW Ari             \\        
HIP~11390   & C & 1999.7728 &  13    & 62.18  &   0.02  &  155.1  &  0.1  &  0.6  &                      \\ 
HIP~11511   & B & 1998.8063 &  22    & 15.63  &   0.02  &  214.6  &  0.1  &  0.6  & A=HIP 11510          \\ 
HIP~11511   & C & 1998.8063 &  14    & 53.40  &   0.01  &  291.1  &  0.1  &  0.6  & New cmp              \\ 
HIP~11572   & B & 1998.7948 &  21    &  5.865 &   0.003 &  231.18 &  0.02 &  2    & C not measured       \\ 
HIP~12781   & - & 2000.8239 &  37    &  -     &   -     &    -    &  -    &  2    & GJ 109 (DMSA/V), no detection \\ 
HIP~15844   & B & 2000.8213 &  23    &  2.397 &   0.003 &  340.90 &  0.05 &  2    & GJ 140 AB, C not in the field \\ 
HIP~17102   & B & 2000.8351 &  48    & 15.445 &   0.004 &  291.30 &  0.01 &  2    & Wo 9119 AB           \\
HIP~17102*  & B & 2004.8842 &  15    & 15.467 &   0.006 &  294.11 &  0.02 &  2    & Wo 9119 AB           \\ 
HIP~17666   & B & 2000.8267 &  48    &  7.139 &   0.003 &   51.81 &  0.02 &  2    & GJ 1064 AB, B=V580 Per\\   
HIP~21088   & B & 2000.8213 &  15    &  9.022 &   0.008 &   61.81 &  0.08 &  2    & GJ 169.1 AB          \\ 
HIP~21765   & B & 2000.8324 &  14    &  -     &   -     &    -    &  -    &  2    & Wo 9163 AB, B not detected \\ 
HIP~22715   & B & 2000.8324 &   5    &  3.980 &   0.003 &  216.32 &  0.03 &  2    & GJ 2035 AB           \\
HIP~22715*  & B & 2004.8843 &  15    &  3.972 &   0.006 &  220.03 &  0.04 &  2    & GJ 2035 AB           \\ 
HIP~24220   & B & 1998.8095 &   6    &  7.07  &   0.07  &  299.6  &  0.6  &  0.6  &                      \\
HIP~24220   & C & 1998.8095 &  30    & 53.74  &   0.01  &  322.5  &  0.1  &  0.6  & New cmp              \\
HIP~24220   & D & 1998.8095 &  29    & 35.32  &   0.06  &  234.2  &  0.1  &  0.6  & New cmp              \\
HIP~24220*  & B & 2004.8819 &  30    &  5.620 &   0.020 &  301.82 &  0.08 &  2    &                      \\ 
HIP~24220*  & C & 2004.8819 &  30    & 53.769 &   0.005 &  322.56 &  0.01 &  2    & New cmp              \\
HIP~24220*  & D & 2004.8819 &  30    & 35.187 &   0.014 &  234.15 &  0.02 &  2    & New cmp              \\
\hline
\end{tabular}
\end{table*}

%Table 4 continued
\setcounter{table}{3}
\begin{table*}[t]
\caption[]{- continued}
\begin{tabular}{llcccccccc}
\hline
Identifier &  Cp   & Epoch             & N\_frames & $\rho$      & $\sigma$($\rho$) & $\theta$   & $\sigma$($\theta$) & Telesc.$^1$ & Remark/     \\
           &       & ($Bessel~yr$) &    & ($\arcsec$) &($\arcsec$)       & ($\degr$ ) & ($\degr$)          &             & Other identifier  \\
\hline
HIP~28368   & - & 2000.8325 & 20   &  -     &   -     &    -     &  -    &  2    & NN 3371 A (DMSA/X), no detection \\ 
HIP~29316   & B & 2000.8216 & 20   &  1.803 &   0.020 &   27.04 &  0.50 &  2     & GJ 228 AB            \\ 
HIP~29316   & C & 2000.8216 & 20   & 12.804 &   0.004 &  124.31 &  0.05 &  2     & New cmp              \\
HIP~30920   & B & 2000.8325 &  0   &   -    &     -   &     -   &   -   &  2     & GJ 234 AB (DMSA/G), B not detected\\
HIP~30920   & C'& 2000.8325 & 44   & 13.677 &   0.016 &   10.38 &  0.01 &  2     & Not a true cmp       \\
HIP~30920   & D'& 2000.8325 & 44   & 25.704 &   0.020 &  158.07 &  0.02 &  2     & Not a true cmp       \\
HIP~31635   & C'& 2000.8298 & 50   & 12.827 &   0.001 &  111.65 &  0.02 &  2     & Not a true cmp       \\
HIP~31635   & D'& 2000.8298 & 48   & 19.365 &   0.003 &  205.88 &  0.01 &  2     & Not a true cmp       \\
HIP~31635   & B & 2000.8298 & 47   & 35.551 &   0.003 &  105.33 &  0.01 &  2     & GJ 239 AB, previously $\rho$=3.7"\\
HIP~34222   & B & 2002.9104 & 12   & 13.006 &   0.003 &  316.85 &  0.01 &  2     & GJ 265 AB            \\
HIP~34222*  & B & 2004.8821 & 30   & 13.081 &   0.006 &  318.89 &  0.02 &  2     & GJ 265 AB            \\ 
HIP~34222   & C'& 2004.8821 & 30   & 74.134 &   0.007 &  125.43 &  0.01 &  2     & Not a true cmp       \\
HIP~34222   & D'& 2004.8821 & 30   & 81.733 &   0.014 &  145.38 &  0.01 &  2     & Not a true cmp       \\
HIP~34222   & E'& 2004.8821 & 30   & 97.571 &   0.012 &  204.77 &  0.01 &  2     & Not a true cmp       \\
HIP~34222   & F'& 2004.8821 & 30   & 95.613 &   0.018 &  218.56 &  0.01 &  2     & Not a true cmp       \\
HIP~39721   & B & 2004.8844 & 15   &  5.042 &   0.001 &  240.10 &  0.01 &  2     & Wo 9251 AB           \\
HIP~39896   & B & 2002.9103 & 12   & 13.710 &   0.004 &  239.31 &  0.01 &  2     & GJ 1108 AB, A=FP Cnc \\
HIP~39896*  & B & 2004.8843 & 15   & 13.750 &   0.002 &  240.32 &  0.02 &  2     & GJ 1108 AB, A=FP Cnc \\ 
HIP~41824   & B & 2000.8353 & 18   & 10.149 &   0.004 &  344.76 &  0.01 &  2     & GJ 2069 AB  (DMSA/V) \\ 
HIP~41824   & C & 2000.8353 & 18   & 21.518 &   0.021 &  304.90 &  0.01 &  2     & New cmp; A=CU Cnc      \\
HIP~43422   & B & 2002.9106 & 12   &  1.717 &   0.001 &  153.07 &  0.04 &  2     & GJ 323 AB (DMSA/X)     \\ 
HIP~44295   & B & 2002.9104 & 12   &  5.119 &   0.002 &  179.87 &  0.02 &  2     & GJ 1120 AB             \\ 
HIP~44295*  & B & 2004.8845 & 15   &  5.100 &   0.003 &  180.80 &  0.01 &  2     & GJ 1120 AB             \\
HIP~54658   & B & 2002.9105 & 12   &  3.736 &   0.001 &   79.56 &  0.01 &  2     &                        \\ 
HIP~67422   & B & 2000.4212 & 27   &  3.30  &   0.01  &  170.7  &  0.1  &  0.6   &                  \\ 
HIP~72659   & B & 2000.4211 & 27   &  6.72  &   0.01  &  314.2  &  0.1  &  0.6   & GJ 566 AB        \\ 
HIP~79607   & B & 2000.4213 & 27   &  6.96  &   0.01  &  232.8  &  0.1  &  0.6   &                  \\ 
HIP~88601   & B & 2000.4214 & 27   &  3.73  &   0.04  &  145.6  &  0.6  &  0.6   &                  \\ 
HIP~92836   & B & 2001.8637 & 48   &  4.021 &   0.007 &   32.92 &  0.02 &  2     & GJ 734 AB        \\ 
HIP~95071   &BQ & 2000.8343 & 19   &  5.030 &   0.015 &  246.57 &  0.14 &  2     & GJ 754.1 B (DMSA/X)\\
HIP~95071   & C & 2000.8343 & 19   & 19.961 &   0.008 &  312.22 &  0.06 &  2     & New cmp          \\
HIP~95071   &BA & 2000.8343 & 19   & 27.152 &   0.005 &  123.26 &  0.11 &  2     & GJ 754.1 BA      \\
HIP~95593   & B & 1998.7918 &  9   &  5.599 &   0.014 &   84.80 &  0.10 &  2     &                  \\ 
HIP~95593   & C & 1998.7918 &  9   & 59.331 &   0.016 &  223.96 &  0.02 &  2     & New cmp          \\ 
HIP~96019   &CD & 1998.7919 &  7   &  5.875 &   0.003 &   50.36 &  0.04 &  2     &                  \\ 
HIP~96019   &CE & 1998.7919 &  7   & 31.654 &   0.014 &   72.16 &  0.01 &  2     &                  \\ 
HIP~96019   &CA & 1998.7919 &  7   & 52.491 &   0.019 &   67.50 &  0.01 &  2     & AB=HIP 96025     \\ 
HIP~96019   &CB & 1998.7919 &  7   & 56.500 &   0.074 &   64.16 &  0.10 &  2     & AB=HIP 96025     \\ 
HIP~97237   & B & 2000.8342 & 19   &  1.418 &   0.006 &   46.98 &  0.21 &  2     & GJ 766, Kui 95   \\ 
HIP~97237   & C & 2000.8342 & 19   & 14.848 &   0.008 &   36.51 &  0.01 &  2     & New cmp          \\
\hline
\end{tabular}
\end{table*}

%Table 4 continued
\setcounter{table}{3}
\begin{table*}[t]
\caption[]{- continued}
\begin{tabular}{llcccccccc}
\hline
Identifier &  Cp   & Epoch             & N\_frames & $\rho$      & $\sigma$($\rho$) & $\theta$   & $\sigma$($\theta$) & Telesc.$^1$ & Remark/     \\
           &       & ($Bessel~yr$) &    & ($\arcsec$) &($\arcsec$)       & ($\degr$ ) & ($\degr$)          &             & Other identifier  \\
\hline
HIP~101150  & B & 2000.8341 &     0   &   -     &   -     &    -    &  -    &  2    & Wo 9697 AB, B not detected\\ 
HIP~101150  & C & 2000.8341 &    20   &  10.310 &   0.004 &  298.38 &  0.02 &  2    &                  \\ 
HIP~101150  & D'& 2000.8341 &    20   &  46.550 &   0.004 &  247.52 &  0.01 &  2    & Not a true cmp   \\
HIP~102518  & B & 1998.7849 &    20   &   6.372 &   0.016 &   37.55 &  0.07 &  2    &                  \\ 
HIP~103822  & B & 1999.7723 &    23   &  18.91  &   0.01  &  299.2  &  0.1  &  0.6  &                  \\ 
HIP~103822  & C & 1999.7723 &    23   &  26.60  &   0.02  &  248.4  &  0.1  &  0.6  &                  \\ 
HIP~104210  & B & 1998.7946 &     5   &   3.510 &   0.017 &   30.58 &  0.13 &  2    &                  \\
HIP~104210  & C & 1998.7946 &     5   &  57.918 &   0.016 &  150.20 &  0.01 &  2    &                  \\
HIP~104210* & C & 1999.7750 &     8   &  58.59  &   0.02  &  150.8  &  0.1  &  0.6  &                  \\
HIP~104210  & D'& 1998.7946 &     5   &  30.689 &   0.022 &    9.52 &  0.04 &  2    & Not a true cmp   \\ 
HIP~104837  & B & 1998.8054 &     7   &   3.91  &   0.20  &  254.3  &  1.1  &  0.6  &                  \\ 
HIP~105421  & B & 1998.7945 &     4   &   5.325 &   0.009 &  307.46 &  0.03 &  2    &                  \\ 
HIP~105421* & B & 2001.8608 &    24   &   5.307 &   0.002 &  303.15 &  0.02 &  2    &                  \\ 
HIP~105747  & B & 1998.8087 &     0   &   -     &   -     &    -    &  -    &  0.6  & B not detected   \\ 
HIP~105747  & C & 1998.8087 &     9   &  26.51  &   0.01  &  241.4  &  0.1  &  0.6  &                  \\
HIP~107554  & B & 1998.7945 &     6   &   3.279 &   0.015 &  201.56 &  0.14 &  2    &                  \\
HIP~108888  & B & 1998.8059 &     5   &   7.32  &   0.13  &  188.6  &  0.3  &  0.6  & V394 Lac         \\ 
HIP~108888  & C & 1998.8059 &    10   &  24.41  &   0.02  &  315.6  &  0.1  &  0.6  & New cmp          \\
HIP~108892  & B & 1998.7946 &     1   &   8.619 &   0.001 &  131.58 &  0.01 &  2    &                  \\ 
HIP~108892* & B & 2001.8610 &    36   &   8.592 &   0.006 &  127.27 &  0.02 &  2    & V378 Lac         \\ 
HIP~108892  & C & 2001.8610 &    36   &  57.513 &   0.006 &   39.97 &  0.01 &  2    & New cmp          \\
HIP~110326  & B & 1998.7946 &    10   &  13.012 &   0.008 &   46.93 &  0.02 &  2    &                  \\ 
HIP~110640  & B & 2001.8582 &    55   &   2.103 &   0.009 &  220.94 &  0.27 &  2    & GJ 857.1 AB      \\
HIP~110640  & C & 2001.8582 &    55   &  71.557 &   0.007 &   47.39 &  0.01 &  2    &                  \\
HIP~111172  & BC& 1998.7945 &     9   &   3.728 &   0.007 &   31.90 &  0.07 &  2    &                  \\ 
HIP~111172  & BD& 1998.7945 &     9   &  16.864 &   0.006 &  116.06 &  0.02 &  2    &                  \\ 
HIP~111172  & BE& 1998.7945 &     9   &  35.488 &   0.013 &   78.63 &  0.02 &  2    & New cmp          \\
HIP~111172* & BC& 1999.7751 &    12   &   3.75  &   0.01  &   32.2  &   0.2 &  0.6  &                  \\ 
HIP~111172* & BD& 1999.7751 &    12   &  16.94  &   0.01  &  115.9  &   0.1 &  0.6  &                  \\ 
HIP~111172* & BE& 1999.7751 &    12   &  35.48  &   0.02  &   78.3  &   0.1 &  0.6  & New cmp          \\
HIP~111172  & BF& 1999.7751 &     8   &  70.25  &   0.04  &   65.2  &  0.1  &  0.6  & New cmp          \\
HIP~111279  & B & 1998.8088 &    15   &  21.94  &   0.01  &  218.7  &  0.1  &  0.6  & A=HIP~111277     \\
HIP~113017  & B & 1998.7947 &    16   &   7.755 &   0.003 &  329.10 &  0.02 &  2    &                  \\ 
HIP~113411  & B & 1998.8061 &    13   &   9.28  &   0.05  &   97.7  &  0.2  &  0.6  &                  \\ 
HIP~113437  & B & 2001.8610 &    24   &   1.515 &   0.001 &  252.96 &  0.09 &  2    &                  \\ 
HIP~113876  & B & 1998.7947 &    21   &   2.689 &   0.007 &  172.66 &  0.07 &  2    &                  \\ 
HIP~113876  & C & 1998.7947 &    21   &  47.829 &   0.013 &  234.21 &  0.01 &  2    & New cmp          \\ 
\hline
\end{tabular}
\end{table*}

%Table 4 continued
\setcounter{table}{3}
\begin{table*}[t]
\caption[]{- continued}
\begin{tabular}{llcccccccc}
\hline
Identifier &  Cp  & Epoch & N\_frames & $\rho$      & $\sigma$($\rho$) & $\theta$   & $\sigma$($\theta$) & Telesc.$^1$ & Remark/    \\
           &      & ($Bessel~yr$) &    & ($\arcsec$) &($\arcsec$)       & ($\degr$ ) & ($\degr$)          &             & Other identifier \\
\hline
HIP~117365        & B & 1998.7947 &    9  &   3.464 &   0.002 &  176.88 &  0.03 &  2    &                \\
HIP~117390        & B & 1998.8090 &   15  &  28.51  &   0.01  &  220.2  &  0.1  &  0.6  &                \\
GJ 1047           & B & 2000.8321 &    0  &   -     &   -     &    -    &  -    &  2    & B not detected \\ 
GJ 1047           & C & 2000.8321 &   24  &  31.025 &   0.004 &  233.25 &  0.04 &  2    &                \\ 
GJ 1047           & D & 2000.8321 &   24  &  47.486 &   0.011 &    9.64 &  0.01 &  2    & New cmp        \\
GJ 1103           & B & 2000.8326 &    0  &   -     &   -     &    -    &  -    &  2    & B not detected \\
GJ 1103           & C'& 2000.8326 &   18  &  27.855 &   0.031 &  176.73 &  0.01 &  2    & Not a true cmp \\
GJ 1103           & D'& 2000.8326 &   18  &  43.361 &   0.007 &  243.98 &  0.04 &  2    & Not a true cmp \\
GJ 1103           & E'& 2000.8326 &   18  &  49.270 &   0.005 &  254.66 &  0.03 &  2    & Not a true cmp \\
GJ 1245           & B & 2000.8344 &   20  &   7.035 &   0.003 &   79.63 &  0.02 &  2    & A=V 1581 Cyg   \\
HD 23713          & B & 2000.8296 &   15  &   1.935 &   0.005 &  126.96 &  0.11 &  2    & Wo~9132=Cou~80 \\ 
HD 218587         & B & 1998.7947 &    8  &   3.073 &   0.003 &  146.21 &  0.06 &  2    & New cmp  \\
HD 218587*        & B & 2000.8238 &   20  &   3.112 &   0.031 &  143.57 &  0.24 &  2    & New cmp  \\
HD 251617         & B & 2000.8324 &   19  &  31.717 &   0.005 &  312.78 &  0.01 &  2    & New cmp  \\
HD 251617         & C & 2000.8324 &   19  &  25.085 &   0.002 &  243.03 &  0.01 &  2    & New cmp  \\ 
BD+24$\degr$692   & B & 2000.8297 &   19  &  32.111 &   0.011 &  203.98 &  0.01 &  2    & New cmp  \\
BD-0$\degr$4073   & B & 1998.7944 &    9  &  72.260 &   0.016 &  330.42 &  0.01 &  2    & New cmp  \\
BD-0$\degr$4073   & C & 1998.7944 &    9  &  78.043 &   0.013 &   77.64 &  0.01 &  2    & New cmp  \\
BD+22$\degr$3800  & B & 1998.7944 &    8  &  21.228 &   0.003 &  266.82 &  0.01 &  2    & New cmp  \\
SA 96 36          & B & 2000.8281 &   45  &   2.364 &   0.017 &  179.32 &  0.15 &  2    & New cmp  \\ 
SA 96 737         & B & 2000.8216 &   20  &  23.325 &   0.013 &  199.41 &  0.01 &  2    & New cmp  \\ 
SA 98 193         & B & 2000.8337 &   34  &  22.615 &   0.006 &  259.10 &  0.02 &  2    & New cmp  \\ 
\hline
\end{tabular}                                                      

$^*$ This flag in col.~1 denotes another epoch for the same target \\
$(^1)$ Instrumentation:\\
\vspace{0.2mm} 2 means {2-m telescope at NAO, Rozhen} \\                               
\vspace{0.2mm} 0.6 means {60-cm telescope at AOB, Belogradchik} \\
\end{table*}

\begin{table*}[t]
\caption[]{\label{t5} Differential photometry of the observed stars}
\begin{tabular}{llccccccccc}
\noalign{\smallskip}
\hline
\noalign{\smallskip}
Identifier & Cp &{\small HJD 2450000.+}& $\Delta$V & $\sigma$($\Delta$V) & $\Delta$B-$\Delta$V & $\sigma$($\Delta$B-$\Delta$V) & $\Delta$V-$\Delta$R & $\sigma$($\Delta$V-$\Delta$R) &
$\Delta$V-$\Delta$I & $\sigma$($\Delta$V-$\Delta$I) \struutdown \\
\hline
HIP~473    & B & 2223.2713 &  0.060 & 0.001 & -0.001 & 0.001 &  0.009 & 0.001 &  0.023 & 0.002 \\         
HIP~473    & C'& 2223.2716 &  4.109 & 0.006 & -0.767 & 0.006 & -0.536 & 0.013 & -0.831 & 0.030 \\        
HIP~473    & E & 2223.2716 &  3.060 & 0.003 & -0.840 & 0.003 & -0.599 & 0.005 & -1.081 & 0.004 \\       
HIP~1397   & B & 1104.3415 &  2.242 & 0.005 &  --    & --    &  0.138 & 0.010 &  0.272 & 0.013 \\       
HIP~1860   & B & 2224.2923 &  2.864 & 0.002 & -0.051 & 0.003 &  0.356 & 0.002 &  0.701 & 0.003 \\       
HIP~3589   & B & 1461.5280 &  1.514 & 0.106 &  --    & --    &  --    & --    & -0.969 & 0.131 \\       
HIP~3589   & C & 1461.5279 &  4.539 & 0.016 &  --    & --    &  0.010 & 0.016 & -0.064 & 0.017 \\       
HIP~3589   & D'& 1461.5279 &  5.130 & 0.055 &  --    & --    &  0.215 & 0.055 &  0.229 & 0.055 \\       
HIP~4258   & B & 2223.2938 &  0.430 & 0.001 &  0.035 & 0.001 &  0.018 & 0.001 &  0.033 & 0.001 \\         
HIP~7495   & B & 2225.2678 &  2.216 & 0.016 &  0.397 & 0.026 &  0.225 & 0.022 &  0.390 & 0.017 \\       
HIP~7495   & C & 2225.2678 &  4.781 & 0.003 &  0.341 & 0.007 &  0.227 & 0.015 &  0.498 & 0.011 \\       
HIP~9275   & B & 1845.4376 &  1.070 & 0.001 & -0.027 & 0.002 &  0.277 & 0.002 &  0.625 & 0.002 \\       
HIP~9488   & B & 2225.4056 &  2.321 & 0.002 &  0.168 & 0.002 &  0.625 & 0.012 &  0.913 & 0.005 \\       
HIP~9867   & B?& 1846.4764 &  3.844 & 0.001 & -0.484 & 0.010 & -0.414 & 0.002 & -0.722 & 0.004 \\       
HIP~9867   & C & 1846.4764 &  3.707 & 0.008 & -0.089 & 0.011 & -0.161 & 0.010 & -0.279 & 0.011 \\       
HIP~9867   & E'& 1846.4764 &  1.934 & 0.001 & -0.392 & 0.011 & -0.394 & 0.002 & -0.731 & 0.002 \\       
HIP~10023  & B & 1104.3527 &  2.201 & 0.005 &  --    & --    &  0.176 & 0.005 &  0.352 & 0.005 \\       
HIP~11390  & B & 1461.5582 &  1.639 & 0.001 &  0.007 & 0.003 &  0.001 & 0.002 &  --    & --    \\        
HIP~11390  & C & 1461.5582 &  5.120 & 0.005 &  0.149 & 0.025 &  0.287 & 0.007 &  --    & --    \\        
HIP~11511  & B & 1108.5385 &  3.106 & 0.003 &  --    & --    &  0.255 & 0.004 &  0.553 & 0.004 \\          
HIP~11572  & B & 1104.3647 &  1.724 & 0.003 &  --    & --    &  0.264 & 0.007 &  0.573 & 0.031 \\          
HIP~15844  & B & 1844.5051 &  1.277 & 0.003 & -0.049 & 0.007 &  0.250 & 0.008 &  0.539 & 0.014 \\          
HIP~17102  & B & 1849.5655 &  3.926 & 0.004 &  0.023 & 0.037 &  --    & --    &  --    & --    \\       
HIP~17102* & B & 3328.4708 &  3.866 & 0.010 &  --    & --    &  0.783 & 0.021 &  1.851 & 0.012 \\       
HIP~17666  & B & 1846.4959 &  0.587 & 0.003 &  0.137 & 0.006 &  0.087 & 0.004 &  0.142 & 0.004 \\       
HIP~21088  & B & 1844.5269 &  1.398 & 0.001 & -1.025 & 0.002 & -1.272 & 0.004 & -2.480 & 0.008 \\       
HIP~22715  & B & 3328.4952 &  4.359 & 0.011 &  --    & --    &  0.718 & 0.014 &  1.540 & 0.014 \\       
HIP~24220  & B & 1109.7223 &  4.835 & 0.059 &  --    & --    &  --    & --    &  --    & --    \\       
HIP~24220  & C & 1109.7308 &  2.046 & 0.010 &  --    & --    &  0.221 & 0.133 &  0.926 & 0.060 \\       
HIP~24220  & D & 1109.7305 &  4.463 & 0.021 &  --    & --    & -0.172 & 0.152 &  0.289 & 0.023 \\       
HIP~24220* & B & 3327.6271 &  3.344 & 0.035 &  --    & --    &  0.071 & 0.039 &  0.126 & 0.039 \\       
HIP~24220* & C & 3327.6271 &  1.991 & 0.002 &  --    & --    &  0.545 & 0.003 &  0.956 & 0.003 \\       
HIP~24220* & D & 3327.6271 &  4.605 & 0.012 &  --    & --    &  0.169 & 0.016 &  0.393 & 0.018 \\          
HIP~29316  & B & 1844.6290 &  1.629 & 0.059 &  0.023 & 0.067 & -0.056 & 0.062 &  0.154 & 0.066 \\       
HIP~29316  & C & 1844.6290 &  4.450 & 0.007 & -0.759 & 0.008 & -0.834 & 0.008 & -1.521 & 0.021 \\       
HIP~30920  & C'& 1848.6091 &  3.857 & 0.002 & -0.064 & 0.004 & -0.810 & 0.004 & -1.585 & 0.010 \\       
HIP~30920  & D'& 1848.6091 &  3.267 & 0.002 & -0.589 & 0.003 & -1.093 & 0.006 & -2.096 & 0.017 \\       
HIP~31635  & C & 1847.6097 &  5.327 & 0.003 &  --    & --    &  --    & --    &  --    & --    \\       
HIP~31635  & D & 1847.6098 &  5.694 & 0.005 &  --    & --    &  --    & --    &  --    & --    \\       
HIP~31635  & E & 1847.6097 &  4.726 & 0.005 &  --    & --    &  --    & --    &  --    & --    \\       
\hline
\end{tabular}
\end{table*}

%Table 5 continued                                                                        
\setcounter{table}{4}                                                                     
\begin{table*}[t]                                                                        
\caption[]{- continued}                                                                   
\begin{tabular}{llccccccccc}
\hline
Identifier & Cp &{\small  HJD 2450000.+}& $\Delta$V & $\sigma$($\Delta$V) & $\Delta$B-$\Delta$V & $\sigma$($\Delta$B-$\Delta$V) & $\Delta$V-$\Delta$R & $\sigma$($\Delta$V-$\Delta$R) &
$\Delta$V-$\Delta$I & $\sigma$($\Delta$V-$\Delta$I) \struutdown \\
\hline
HIP~34222   & B  & 2607.5388 &  4.380 & 0.006 & -0.240 & 0.010 & -0.270 & 0.008 & -0.437 & 0.026 \\       
HIP~34222   & B  & 3327.6817 &  4.410 & 0.009 &  --    & --    & -0.226 & 0.015 & -0.318 & 0.024 \\       
HIP~34222   & C  & 3327.6817 &  4.027 & 0.005 &  --    & --    & -0.493 & 0.014 & -0.839 & 0.021 \\       
HIP~34222   & D  & 3327.6817 &  4.357 & 0.004 &  --    & --    & -0.486 & 0.017 & -0.842 & 0.048 \\       
HIP~34222   & E  & 3327.6817 &  2.741 & 0.002 &  --    & --    & -0.436 & 0.006 & -0.735 & 0.011 \\       
HIP~34222   & F  & 3327.6817 &  4.627 & 0.005 &  --    & --    & -0.547 & 0.022 & -0.937 & 0.043 \\      
HIP~39721   & B  & 3328.5223 &  0.282 & 0.004 &  --    & --    & -0.127 & 0.005 & -0.229 & 0.005 \\       
HIP~39896   & B  & 2607.5013 &  2.066 & 0.007 &  0.075 & 0.009 &  0.397 & 0.020 &  0.964 & 0.020 \\      
HIP~39896   & B  & 3328.5068 &  1.101 & 0.007 &  --    & --    & -0.524 & 0.011 & -1.010 & 0.008 \\         
HIP~41824   & B  & 1849.6093 &  1.458 & 0.004 & -0.094 & 0.005 &  0.155 & 0.006 &  0.208 & 0.008 \\      
HIP~41824   & C  & 1849.6093 &  3.872 & 0.003 & -0.602 & 0.005 & -1.013 & 0.008 & -2.030 & 0.009 \\      
HIP~43422   & B  & 2607.6096 &  0.305 & 0.003 &  0.042 & 0.009 &  0.052 & 0.014 &  0.114 & 0.004 \\      
HIP~44295   & B  & 2607.5365 &  0.204 & 0.010 &  0.064 & 0.010 &  0.025 & 0.012 &  0.079 & 0.014 \\      
HIP~44295   & B  & 3328.5448 &  0.205 & 0.001 &  --    & --    &  0.073 & 0.003 &  0.035 & 0.004 \\      
HIP~54658   & B  & 2607.5714 &  0.114 & 0.003 & -0.010 & 0.004 &  0.023 & 0.004 &  0.053 & 0.005 \\
HIP~67422   & B  & 1698.3758 &  0.31  & 0.02  &   --   & --    &  0.06  & 0.02  &  0.15  & 0.02  \\      
HIP~72659   & B  & 1698.3392 &  2.24  & 0.01  &   --   & --    &  0.32  & 0.01  &  0.60  & 0.02  \\      
HIP~79607   & B  & 1698.4123 &  0.88  & 0.01  &   --   & --    &  0.01  & 0.02  & -0.06  & 0.01  \\      
HIP~88601   & B  & 1698.4489 &  1.67  & 0.05  &   --   & --    &  0.28  & 0.10  &  0.48  & 0.08  \\      
HIP~92836   & B  & 2225.2124 &  3.304 & 0.011 &  0.061 & 0.011 &  0.518 & 0.013 &  1.054 & 0.060 \\      
HIP~95071   &BQ  & 1849.2403 &  0.619 & 0.001 & -0.161 & 0.004 &  --    & --    &  --    & --    \\      
HIP~95071   & C  & 1849.2403 &  4.017 & 0.006 & -0.239 & 0.008 &  --    & --    &  --    & --    \\      
HIP~95071   &BA  & 1849.2403 &  0.242 & 0.006 & -1.415 & 0.006 &  --    & --    &  --    & --    \\      
HIP~95593   & B  & 1103.2554 &  0.368 & 0.018 &  --    & --    & -0.021 & 0.018 & -0.064 & 0.029 \\      
HIP~95593   & C  & 1103.2554 &  5.890 & 0.011 &  --    & --    &  1.733 & 0.011 &  3.758 & 0.013 \\      
HIP~96019   &CD  & 1103.2866 &  0.366 & 0.004 &  --    & --    &  0.028 & 0.004 &  0.057 & 0.006 \\      
HIP~96019   &CE  & 1103.2866 &  4.123 & 0.003 &  --    & --    &  0.454 & 0.003 &  0.899 & 0.005 \\      
HIP~96019   &CA  & 1103.2866 &  0.037 & 0.002 &  --    & --    & -0.009 & 0.002 & -0.004 & 0.004 \\      
HIP~96019   &CB  & 1103.2866 &  3.620 & 0.037 &  --    & --    & -0.031 & 0.037 &  0.048 & 0.095 \\      
HIP~97237   & B  & 1849.2115 &  0.609 & 0.013 &  0.030 & 0.028 &  0.161 & 0.044 &  0.316 & 0.074 \\      
HIP~97237   & C  & 1849.2115 &  3.447 & 0.006 &  0.343 & 0.011 & -0.288 & 0.016 & -0.638 & 0.034 \\      
HIP~101150  & C  & 1849.1984 &  3.324 & 0.004 & -0.673 & 0.006 &  --    & --    & -0.830 & 0.005 \\      
HIP~101150  & D' & 1849.1984 &  3.608 & 0.003 & -0.750 & 0.006 &  --    & --    & -0.864 & 0.006 \\      
HIP~102518  & B  & 1100.7335 &  3.637 & 0.007 &  --    & --    &  0.196 & 0.009 &  0.508 & 0.020 \\      
HIP~103822  & B  & 1461.3837 &  3.282 & 0.003 &  --    & --    & -0.666 & 0.004 & -1.356 & 0.009 \\      
HIP~103822  & C  & 1461.3837 &  7.212 & 0.029 &  --    & --    &  1.482 & 0.030 &  3.307 & 0.030 \\      
HIP~104210  & B  & 1104.2639 &  0.709 & 0.003 &  --    & --    &  0.084 & 0.003 &  0.229 & 0.009 \\      
HIP~104210  & C  & 1104.2639 &  3.089 & 0.028 &  --    & --    &  0.804 & 0.028 &  1.742 & 0.030 \\      
HIP~104210* & C  & 1462.3546 &  3.463 & 0.004 &  --    & --    &  0.755 & 0.008 &  1.693 & 0.006 \\      
HIP~104210  & D' & 1104.2639 &  5.222 & 0.034 &  --    & --    &  0.301 & 0.034 &  0.510 & 0.039 \\      
\hline
\end{tabular}
\end{table*}

%Table 5 continued                                                                        
\setcounter{table}{4}                                                                     
\begin{table*}[t]                                                                        
\caption[]{- continued}                                                                   
\begin{tabular}{llccccccccc}
\hline
Identifier & Cp &{\small  HJD 2450000.+}& $\Delta$V & $\sigma$($\Delta$V) & $\Delta$B-$\Delta$V & $\sigma$($\Delta$B-$\Delta$V) & $\Delta$V-$\Delta$R & $\sigma$($\Delta$V-$\Delta$R) &
$\Delta$V-$\Delta$I & $\sigma$($\Delta$V-$\Delta$I) \struutdown \\
\hline
HIP~104837  & B  & 1108.2303 &  1.331 & 0.153 &  --    & --    &  0.139 & 0.163 &  --    & --    \\      
HIP~105421  & B  & 2224.1943 &  3.919 & 0.002 & 0.738  & 0.005 & (0.432)&(0.016)&  0.766 & 0.004 \\      
HIP~105747  & C  & 1109.4354 &  2.062 & 0.002 &  --    & --    &  0.120 & 0.005 &  0.247 & 0.002 \\      
HIP~107554  & B  & 1104.2359 &  2.852 & 0.001 &  --    & --    &  0.074 & 0.053 &  0.241 & 0.031 \\      
HIP~108888  & B  & 1108.3842 &  3.173 & 0.023 &  --    & --    &  --    & --    &  --    & --    \\      
HIP~108892  & B  & 1104.2690 &  2.960 & 0.001 &  --    & --    &  --    & --    &  --    & --    \\      
HIP~108892* & B  & 2224.2363 &  3.118 & 0.110 & -1.166 & 0.110 & -0.742 & 0.110 & -1.295 & 0.111 \\      
HIP~108892  & C  & 2224.2363 &  3.346 & 0.088 & -0.773 & 0.089 & -0.582 & 0.089 & -1.024 & 0.090 \\      
HIP~110326  & B  & 1104.2870 &  1.766 & 0.014 &  --    & --    & -0.260 & 0.031 & -0.096 & 0.014 \\      
HIP~110640  & B  & 2223.2229 &  2.378 & 0.028 &  0.319 & 0.079 &  0.371 & 0.030 &  0.814 & 0.033 \\      
HIP~110640  & C  & 2223.2229 &  4.277 & 0.003 & -0.655 & 0.006 & -0.376 & 0.005 & -0.620 & 0.007 \\      
HIP~111172  & BC & 1104.2439 &  0.902 & 0.005 &  --    & --    &  0.085 & 0.007 &  0.140 & 0.011 \\      
HIP~111172  & BD & 1104.2439 &  4.125 & 0.015 &  --    & --    &  0.219 & 0.017 &  0.452 & 0.033 \\      
HIP~111172  & BE & 1104.2439 &  5.063 & 0.020 &  --    & --    & -0.011 & 0.029 & -0.069 & 0.028 \\      
HIP~111172* & BC & 1462.3869 &  0.979 & 0.004 &  --    & --    & -0.009 & 0.016 &  0.098 & 0.016 \\      
HIP~111172* & BD & 1462.3869 &  4.042 & 0.006 &  --    & --    &  0.173 & 0.011 &  0.516 & 0.014 \\      
HIP~111172* & BE & 1462.3869 &  5.072 & 0.011 &  --    & --    & -0.059 & 0.036 & -0.053 & 0.012 \\      
HIP~111172* & BC & 1462.3833 &  0.946 & 0.041 &  --    & --    &  0.031 & 0.101 &  0.116 & 0.077 \\      
HIP~111279  & B  & 1109.4570 &  2.789 & 0.001 &  --    & --    &  0.021 & 0.006 &  0.106 & 0.002 \\     
HIP~113017  & B  & 1104.2978 &  2.594 & 0.003 &  --    & --    &  0.331 & 0.006 &  0.713 & 0.006 \\     
HIP~113411  & B  & 1108.4828 &  2.517 & 0.021 &  --    & --    &  0.147 & 0.022 &  0.424 & 0.021 \\     
HIP~113437  & B  & 2224.2609 &  0.355 & 0.011 &  0.043 & 0.014 &  0.019 & 0.024 &  0.051 & 0.013 \\     
HIP~113876  & B  & 1104.3094 &  2.187 & 0.013 &  --    & --    &  0.201 & 0.031 &  0.340 & 0.016 \\     
HIP~113876  & C  & 1104.3094 &  5.664 & 0.007 &  --    & --    &  0.184 & 0.014 &  0.273 & 0.011 \\     
HIP~117365  & B  & 1104.3199 &  0.127 & 0.006 &  --    & --    &  0.025 & 0.008 &  0.033 & 0.008 \\     
HIP~117390  & B  & 1109.5318 &  2.087 & 0.002 &  --    & --    & -0.128 & 0.002 &  --    & --    \\      
GJ 1047     & C  & 1848.4714 &  0.809 & 0.011 & -0.045 & 0.081 &  0.017 & 0.016 &  0.057 & 0.012 \\         
GJ 1047     & D  & 1848.4714 &  1.972 & 0.191 & -0.601 & 0.213 & -0.710 & 0.191 & -1.632 & 0.191 \\         
GJ 1103     & C' & 1848.6455 &  2.594 & 0.005 & -0.489 & 0.008 & -1.153 & 0.010 & -2.239 & 0.012 \\         
GJ 1103     & D' & 1848.6455 &  4.219 & 0.003 & -0.492 & 0.007 & -1.096 & 0.012 & -2.178 & 0.008 \\         
GJ 1103     & E' & 1848.6455 &  3.910 & 0.004 & -0.241 & 0.006 & -0.911 & 0.012 & -1.839 & 0.008 \\         
GJ 1245     & B  & 1849.2899 &  0.627 & 0.004 & -0.077 & 0.002 &  0.141 & 0.004 &  0.152 & 0.004 \\         
HD 23713    & B  & 1847.5296 &  1.190 & 0.015 &  0.211 & 0.016 &  0.126 & 0.032 &  0.127 & 0.035 \\
HD 218587   & B  & 1104.3270 &  3.076 & 0.013 &  --    & --    &  0.364 & 0.014 &  0.574 & 0.032 \\         
HD 218587*  & B  & 1845.4075 &  3.032 & 0.071 &  0.533 & 0.075 &  0.339 & 0.074 &  0.569 & 0.075 \\         
HD 251617   & B  & 1848.5585 &  2.728 & 0.006 &  0.859 & 0.006 &  0.542 & 0.017 &  1.018 & 0.008 \\       
HD 251617   & C  & 1848.5585 &  3.885 & 0.004 &  1.140 & 0.007 &  0.658 & 0.005 &  1.223 & 0.006 \\         
BD+24$\degr$692 & B & 1847.5639 & 3.857 & 0.004 &  0.614 & 0.005 &  --    & --    &  --    & --    \\         
BD+40$\degr$73  & B & 1104.2052 & 2.283 & 0.004 &  --    & --    &  0.081 & 0.007 &  0.139 & 0.005 \\         
BD+40$\degr$73  & C & 1104.2052 & 5.404 & 0.008 &  --    & --    & -0.108 & 0.009 & -0.192 & 0.009 \\         
\hline
\end{tabular}
\end{table*}

%Table 5 continued
\setcounter{table}{4}
\begin{table*}[t]
\caption[]{- continued}
\begin{tabular}{llccccccccc}
\hline
Identifier &  Cp  &{\small  HJD 2450000.+}& $\Delta$V & $\sigma$($\Delta$V) & $\Delta$B-$\Delta$V & $\sigma$($\Delta$B-$\Delta$V) & $\Delta$V-$\Delta$R & $\sigma$($\Delta$V-$\Delta$R) &
$\Delta$V-$\Delta$I & $\sigma$($\Delta$V-$\Delta$I) \struutdown \\
\hline
BD+22$\degr$3800 & B & 1104.1939 & 4.244 & 0.003 &  --    & --    &  0.167 & 0.004 &  0.336 & 0.006 \\         
SA 96 36   & B & 1846.9934 & 2.925 & 0.032 &  0.481 & 0.046 &  0.245 & 0.054 &  0.564 & 0.037 \\        
SA 96 737  & B & 1844.6131 & 0.241 & 0.001 & -0.822 & 0.002 & -0.488 & 0.003 & -0.874 & 0.002 \\
SA 98 193  & B & 1849.0464 & 0.566 & 0.012 & -0.948 & 0.014 & -0.504 & 0.015 & -0.901 & 0.017 \\
\hline
\end{tabular}

$^*$ This flag in col.~1 denotes another epoch for the same target \\
\end{table*}

\begin{table*}[t]
\caption[]{\label{t6} Relative position compared to the Hipparcos/CCDM data}
\begin{tabular}{llcccccccc}
\noalign{\smallskip}
\hline
\noalign{\smallskip}
Identifier       & Cp &  N\_frames & Epoch  & $\rho(\arcsec)$ & $\theta(\degr)$ & $\delta\rho(\arcsec)$ & $\delta\theta(\degr)$ & $\Delta$Pos($\arcsec$) & Code \struutdown \\
\hline
HIP~473          & B  &  24   & 1991.25 &  6.041 &  178.25 &  0.038 &   0.240 &  0.046 &  O (cf. Tab~7)                               \\
HIP~1397         & B  &  26   & 1991.25 &  7.519 &  224.61 &  0.849 &  -2.940 &  0.942 &  L (high pm)                                 \\
HIP~1860         & B  &  58   & 1965    & 11.2   &   63.3  & -0.022 &  -5.165 &  1.009 &  M (high pm; CPM)                            \\
HIP~3589         & B  &  12   & 1991.25 &  3.802 &   90.3  & -0.705 &  -6.962 &  0.819 &  L (high pm)                                 \\
HIP~4258         & B  &  24   & 1991.25 &  6.486 &   71.   &  0.002 &  -4.110 &  0.465 &  M (high pm; orbital?)                       \\
HIP~7495         & B  &  55   & 1991.25 &  1.86  &**309.0  & -0.001 &  -5.164 &  0.168 &  S                                           \\
HIP~8414         & C  &   2   & 1878    & 53.4   & *146.   & -0.228 &  -3.284 &  3.062 &  S                                           \\
HIP~9275         & B  &  40   & 1991.25 &  3.798 &   59.4  &  0.128 &  -4.826 &  0.350 &  M (high pm)                                 \\
HIP~9488         & B  &  24   & 1944    & 17.6   &   14.   &  0.167 &  -2.850 &  0.895 &  S                                           \\
HIP~9867         & B? &  52   & 2001    & 26.1   &  326.   & -0.749 &  -2.487 &  1.344 &  L (high pm; probably B)                     \\
HIP~9867         & C  &  52   & 2001    & 10.    &  238.   & -0.004 &  -7.107 &  1.239 &  L (high pm)                                 \\
HIP~10023        & B  &   5   & 1991.25 &  6.173 &  254.4  &  0.026 &   0.130 &  0.030 &  S                                           \\
HIP~11390        & B  &  13   & 1875    & 73.8   &   31.   &  0.200 &   0.700 &  0.925 &  S                                           \\        
HIP~11390        & C  &  13   & 1898    & 62.3   &  155.   & -0.120 &   0.100 &  0.162 &  S                                           \\ 
HIP~11511        & B  &  22   & 1991.25 & 15.74  &  215.0  & -0.110 &  -0.406 &  0.156 &  S                                           \\
HIP~11572        & B  &  21   & 1991.25 &  5.814 &  230.9  &  0.051 &   0.280 &  0.058 &  S                                           \\
HIP~15844        & B  &  23   & 1991.25 &  2.247 &  347.1  &  0.150 &  -6.195 &  0.292 &  M (high pm; Pap~II)                         \\
HIP~17102*       & B  &  15   & 1991.25 & 14.2   &  294.   &  1.267 &   0.112 &  1.268 &  S ?                                         \\
HIP~17666        & B  &  48   & 1991.25 &  7.307 &   54.17 & -0.168 &  -2.356 &  0.341 &  M (high pm; CPM; Pap~II)                    \\
HIP~21088        & B  &  15   & 1991.25 &  8.350 &   69.4  &  0.672 &  -7.593 &  1.331 &  M (high pm; CPM; Pap~II)                    \\
HIP~22715        & B  &  15   & 1901    &  4.4   &  201.   & -0.428 &  19.032 &  1.447 &  M (orbital ?)                               \\
HIP~24220        & B  &  30   & 1991.25 &  5.656 &  301.7  & -0.036 &   0.120 &  0.038 &  S                                           \\
HIP~29316        & B  &  20   & 1991.25 &  2.537 &   25.   & -0.734 &   2.039 &  0.738 &  M (high pm; binary)                         \\
HIP~31635        & B  &  47   & 1962    &  3.7   &   77.   & 31.851 &  28.330 & 32.342 &  L (high pm)                                 \\
HIP~34222*       & B  &  30   & 1959    & 12.1   &  299.   &  0.981 &  19.892 &  4.455 &  L (high dm)                                 \\
HIP~39721        & B  &  15   & 1991.25 &  5.163 &  239.4  & -0.121 &   0.703 &  0.136 &  M ? (high pm)                               \\
HIP~39896*       & B  &  15   & 1960    & 13.3   &  242.   &  0.450 &  -1.680 &  0.600 &  S                                           \\
HIP~41824        & B  &  18   & 1936    & 12.    &  348.   & -1.851 &  -3.238 &  1.953 &  M (high pm; CPM)                            \\ 
HIP~43422        & B  &  12   & 1965    &  2.6   &  121.   & -0.883 &  34.070 &  1.521 &  M (high pm; orbital ?)                      \\
HIP~44295*       & B  &  15   & 1991.25 &  5.169 &  183.3  & -0.069 &  -2.498 &  0.234 &  M (high pm; binary)                         \\
HIP~54658        & B  &  12   & 1991.25 &  3.794 &  79.7   & -0.058 &  -0.140 &  0.059 &  S                                           \\
HIP~92836        & B  &  48   & 1946    &  5.2   &  14.    & -1.179 &  18.925 &  1.911 &  M (high pm; orbital ?)                      \\
HIP~95071        &BQ  &  19   & 1951    &  4.9   &   7.    &  0.330 &-119.432 &  8.404 &  M (high pm; binary)                         \\
HIP~95071        &BA  &  19   & 1892    & 27.5   &  310.   & -0.347 &  -6.740 &  3.231 &  S (high pm; CPM)                            \\
HIP~95593        & B  &   9   & 1991.25 &  5.622 &  85.92  & -0.023 &  -1.120 &  0.112 &  S                                           \\
HIP~96019        &CD  &   7   & 1991.25 &  5.88  &  50.    & -0.005 &   0.360 &  0.037 &  S                                           \\
HIP~97237        & B  &  19   & 1960    &  0.9   & *67.    &  0.518 & -20.022 &  0.650 &  O (cf. Tab~7)                               \\
HIP~101150       & C  &  20   & 1959    & 15.    & 273.    & -4.690 &  25.382 &  7.201 &  L (high pm)                                 \\
HIP~102518       & B  &  20   & 1991.25 &  6.412 &  36.5   & -0.040 &   1.050 &  0.124 &  S                                           \\
\hline
\end{tabular}
\end{table*}

%Table 6 continued
\setcounter{table}{5}
\begin{table*}[t]
\caption[]{- continued}
\begin{tabular}{llcccccccc}
\noalign{\smallskip}
\hline
\noalign{\smallskip}
Identifier       & Cp &  N\_frames & Epoch  & $\rho(\arcsec)$ & $\theta(\degr)$  & $\delta\rho(\arcsec)$ & $\delta\theta(\degr)$ & $\Delta$Pos($\arcsec$) & Code \struutdown \\
\hline
HIP~103822       & B  &  23   & 1991.25 & 18.83  & 299.3   &  0.083 &  -0.103 &  0.089 &  S                                           \\
HIP~103822       & C  &  23   & 1878    & 25.9   & 250.    &  0.704 &  -1.605 &  1.018 &  S                                           \\
HIP~104210       & B  &   5   & 1991.25 &  2.786 &  27.9   &  0.724 &   2.680 &  0.739 &  L                                           \\
HIP~104210       & C  &   5   & 1903    & 60.    & 152.    & -2.082 &  -1.800 &  2.786 &  L                                           \\
HIP~105421       & B  &   4   & 1991.25 &  5.39  & 308.    & -0.065 &  -0.540 &  0.082 &  S                                           \\
HIP~105747       & C  &   9   & 1991.25 & 26.55  & 241.34  & -0.044 &   0.034 &  0.047 &  S                                           \\
HIP~107554       & B  &   6   & 1991.25 &  3.298 & 201.0   & -0.019 &   0.560 &  0.037 &  S                                           \\
HIP~108888       & B  &   5   & 1991.25 &  8.531 & 189.6   & -1.210 &  -1.032 &  1.218 &  S ?                                         \\ 
HIP~108892       & B  &   1   & 1991.25 &  8.485 & 131.2   &  0.134 &   0.380 &  0.146 &  S ?                                         \\
HIP~108892*      & B  &  36   & 1991.25 &  8.485 & 131.2   &  0.107 &  -3.925 &  0.595 &  S ?                                         \\
HIP~110326       & B  &  10   & 1991.25 & 12.589 &  46.10  &  0.423 &   0.830 &  0.462 &  L                                           \\
HIP~110640       & B  &  55   & 1991.25 & 1.613  & 228.    &  0.490 &  -7.059 &  0.540 &  O (high pm; Pap~II)                         \\
HIP~110640       & C  &  55   & 1895    & 45.5   &  51.    & 26.057 &  -3.613 & 26.304 &  L (high pm)                                 \\
HIP~111172       & BC &   9   & 1991.25 &  3.682 &  32.2   &  0.046 &  -0.3   &  0.050 &  S                                           \\
HIP~111172       & BD &   9   & 1892    & 16.6   & 118.    &  0.264 &  -1.940 &  0.625 &  S                                           \\
HIP~111279       & B  &  15   & 1991.25 & 21.351 & 216.3   &  0.593 &   2.431 &  1.093 &  S ?                                         \\
HIP~113017       & B  &  16   & 1991.25 &  7.766 & 328.8   & -0.011 &   0.3   &  0.042 &  S                                           \\
HIP~113411       & B  &  13   & 1991.25 &  9.103 &  96.1   &  0.182 &   1.607 &  0.316 &  S ?                                         \\
HIP~113437       & B  &  24   & 1991.25 &  1.545 & 257.6   & -0.030 &  -4.640 &  0.127 &  M (orbital ?)                               \\
HIP~113876       & B  &  21   & 1991.25 &  2.678 & 172.0   &  0.011 &   0.660 &  0.033 &  S                                           \\
HIP~117365       & B  &   9   & 1991.25 &  3.446 & 177.    &  0.018 &  -0.120 &  0.019 &  S                                           \\
HD~23713         & B  &  15   & 1966    &  0.5   & 109.    &  1.435 &  17.959 &  1.468 &  M (orbital ?)                               \\ 
GJ~1047          & C  &  24   & 1959    & 34.    & 233.    & -2.975 &   0.248 &  2.978 &  M (high pm; CPM)                            \\         
GJ~1245          & B  &  20   & 1954    &  7.9   & 106.    & -0.865 & -26.370 &  3.509 &  M (high pm; CPM)                            \\         
\hline
\end{tabular}

Epoch: 1991.25 from Hipparcos else epoch from CCDM \\              
Code: L=optical; M=motion, O=orbital; S=stable; *=180\degr converted; **=alternative Hipparcos solution\\
$^*$ This flag in col.~1 denotes another epoch for the same target \\
\end{table*}

\end{document}